\documentclass[]{aa}  
\usepackage{graphicx}
\begin{document}
   \title{NGC 3934: a shell galaxy in a compact galaxy environment}


   \author{D. Bettoni
          \inst{1}
          \and G. Galletta\inst{2} \and R. Rampazzo\inst{1}  \and A. Marino \inst{1} \and P. Mazzei \inst{1} \and L. M. Buson\inst{1}}

   \offprints{D. Bettoni}

   \institute{(1) INAF, Osservatorio Astronomico di Padova, Vicolo dell'Osservatorio 5, 35122 Padova, Italy \\
    \email{daniela.bettoni@oapd.inaf.it, antonietta.marino@oapd.inaf.it, roberto.rampazzo@oapd.inaf.it,  \\ paola.mazzei@oapd.inaf.it, lucio.buson@oapd.inaf.it} \\
             (2) Dipartimento di Astronomia, Universit\'a di Padova, Vicolo dell'Osservatorio  3, 35122 Padova, Italy\\
             \email{giuseppe.galletta@unipd.it}
             }

   \date{Received ; accepted }

  \abstract
   {  Mergers/accretions are considered the main drivers of the evolution of 
   galaxies in groups. We investigate the NGC~3933 poor galaxy association,
   that contains NGC~3934, which is classified as a polar-ring galaxy.}
   {The multi-band photometric analysis of NGC 3934 allows us to investigate the nature of this galaxy and to re-define the NGC~3933 group members with the aim to characterize the group dynamical properties and its evolutionary phase.}
   {We imaged the group in the far (FUV, $\lambda_{eff}$ = 1539 \AA) and near  
 (NUV, $\lambda_{eff}$ = 2316 \AA) ultraviolet (UV) bands of the Galaxy Evolution Explorer ({\it GALEX}).
   From the deep optical imaging we determined the fine structure of NGC~3934. 
   We measured the recession velocity of PGC~213894 which shows that it belongs to the
   NGC~3933 group.  We derived the spectral  energy distribution (SED) from FUV ({\it GALEX}) to far-IR emission of the two brightest members of the group.  We compared a grid of smooth particle hydrodynamical (SPH) chemo-photometric simulations with  the SED and the integrated properties of NGC~3934 and NGC~3933 to devise their possible formation/evolutionary scenarios. }
   {The NGC~3933 group has six bright members: a core composed of five galaxies, 
   which have Hickson's compact group characteristics, and a more distant member, PGC~37112. 
   The group velocity dispersion is  relatively low (157$\pm$44 km~s$^{-1}$). The
    projected mass, from the NUV photometry, is $\sim$7$\times$10$^{12}$ M$_\odot$ with
    a crossing time of 0.04 Hubble times, suggesting that at least in the center the group is 
    virialized. We do not find evidence that NGC 3934 is a polar-ring galaxy, as suggested by the literature, 
   but find that it is a disk galaxy with a prominent dust-lane structure and a wide type-II shell structure.
    }
   {  NGC~3934 is a quite rare example of a shell galaxy in a likely dense galaxy region.
   The comparison between physically motivated SPH simulations with multi-band
   integrated photometry suggests that NGC~3934 is the product of a major merger.}


   \maketitle

\section{Introduction}

Owing to the low galaxy velocity dispersion, accretions and major mergers are 
considered the main drivers of the evolution of galaxies that inhabit low-density environments, such as small groups.  Ram-pressure stripping 
(Guun \& Gott 1972; Bekki 2009 and references therein) and galaxy 
``harassment'' (see e.g. Moore et al. 1998) are the type of interactions 
that are thought to dominate within the core of dense clusters. 
The combination of the above type of interactions, certainly evolving with time, is at the
 origin of the segregation of galaxy morphology with radius, which has often been observed
in clusters (Oemler 1974; Dressler 1980). In galaxy groups, also in more dense associations such as compact group of galaxies (CG), observations show a plethora of morphological 
properties (see e.g. Hickson 1982). Some CGs are composed only 
of spiral members; others, characterized by a low level of activity of their members,
are dominated by E/S0 galaxies (ETGs hereafter) with old stellar populations 
(see e.g. Coziol et al. 2000). If interpreted in a hierarchical scenario
as an evolutionary sequence, CGs would evolve from a more active (SF, AGN etc.) 
phase dominated by a population of late-type galaxies to a more relaxed and passive 
phase inhabited by ETGs.   

   \begin{figure*}
   {\includegraphics[width=9.5cm]{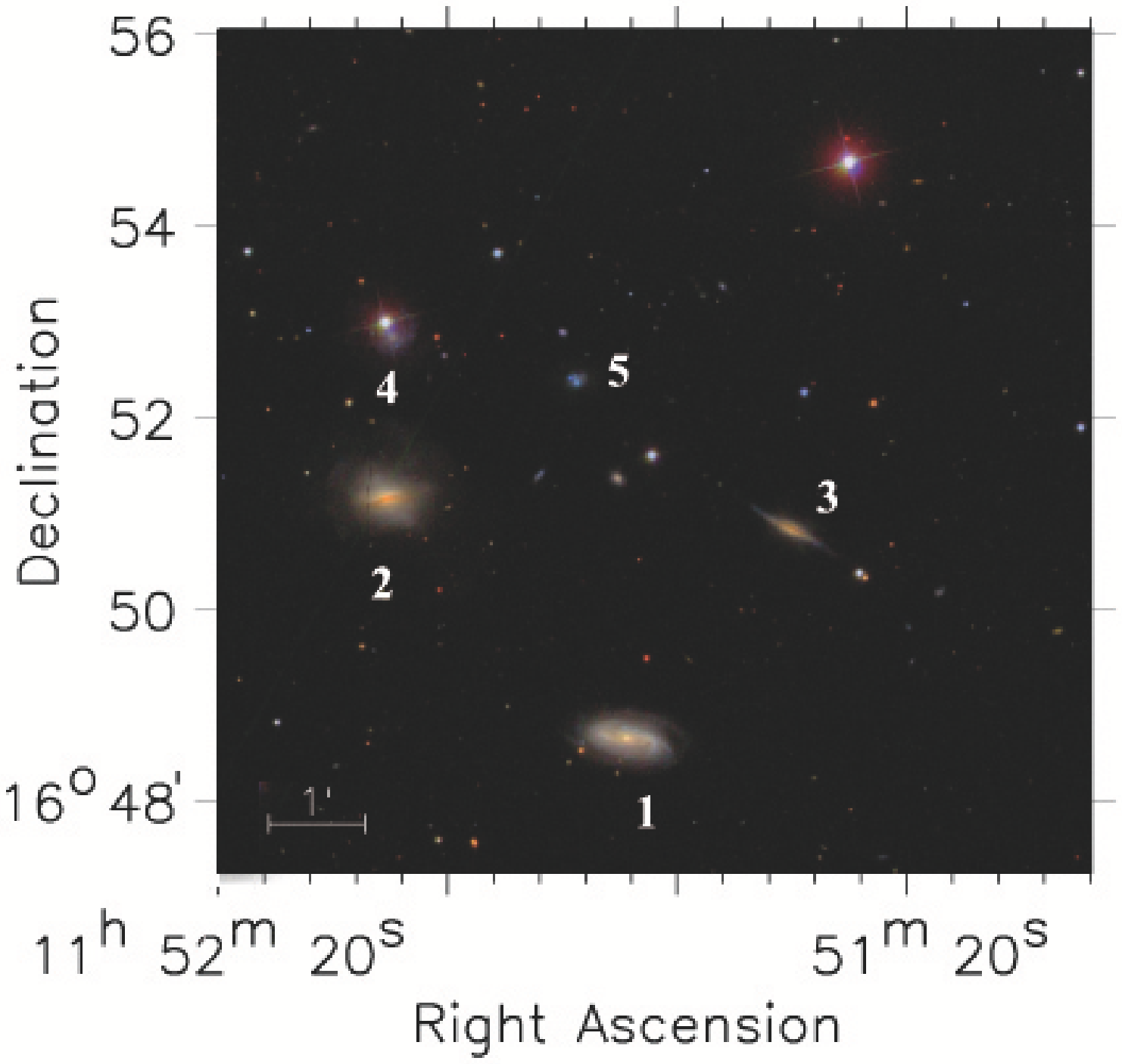} 
   \includegraphics[width=9.5cm]{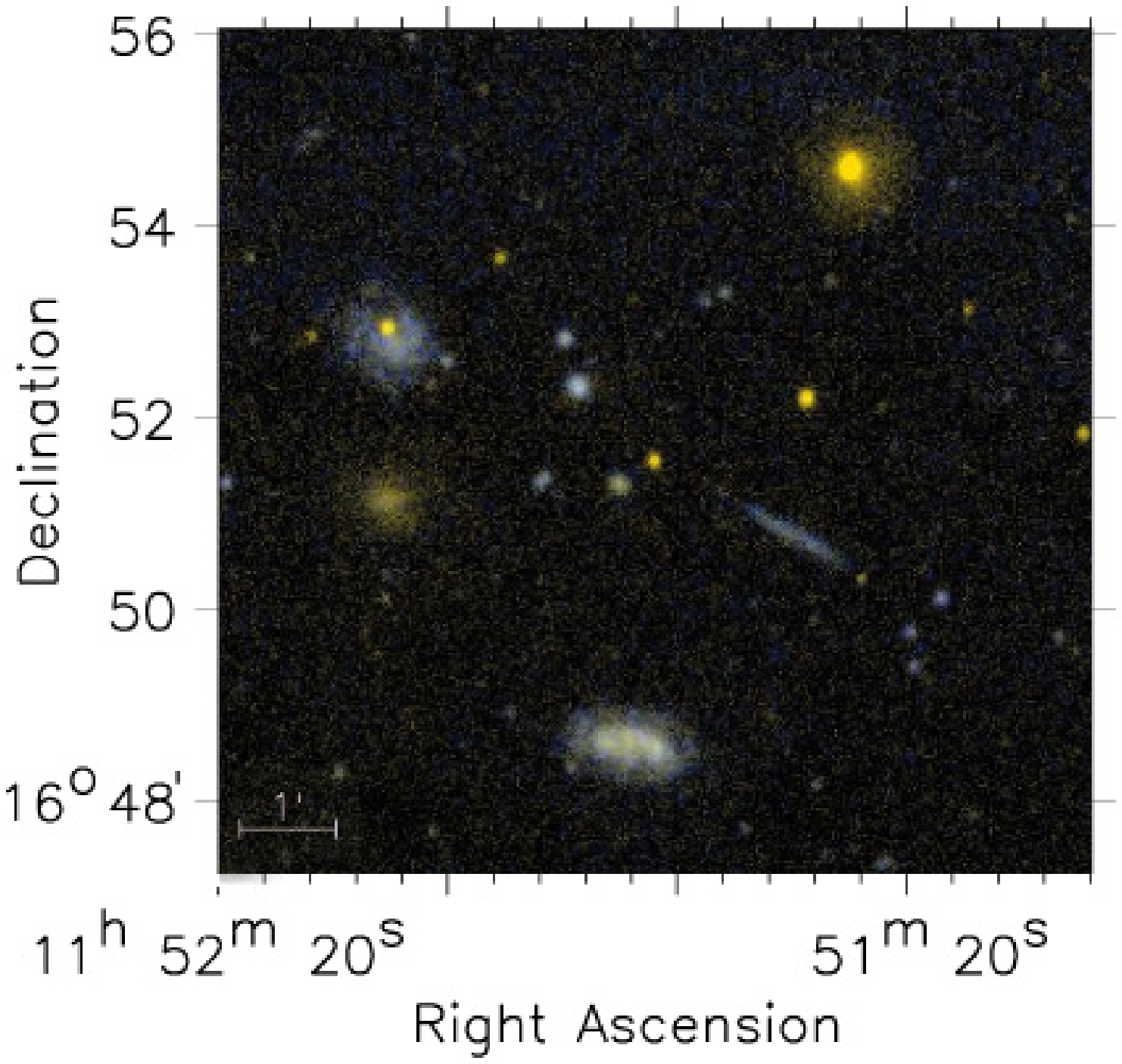}}
      \caption{the NGC 3933 group. ({\it Left panel}) Composite  $u$ (blue), 
      $g$ (yellow), $r$ (red) SDSS image.  
       {\it Right panel}: Full resolution, composite  {\it GALEX} NUV (yellow) and FUV 
       (blue) image. Galaxy members of the group are labeled as in Table~1. The galaxy
       labeled 6 is outside the field because it is ~10.72\arcmin\  southwest of NGC~3933.}
         \label{Fig1}
   \end{figure*}
   \begin{figure*}
   \centering
   \includegraphics[width=\textwidth]{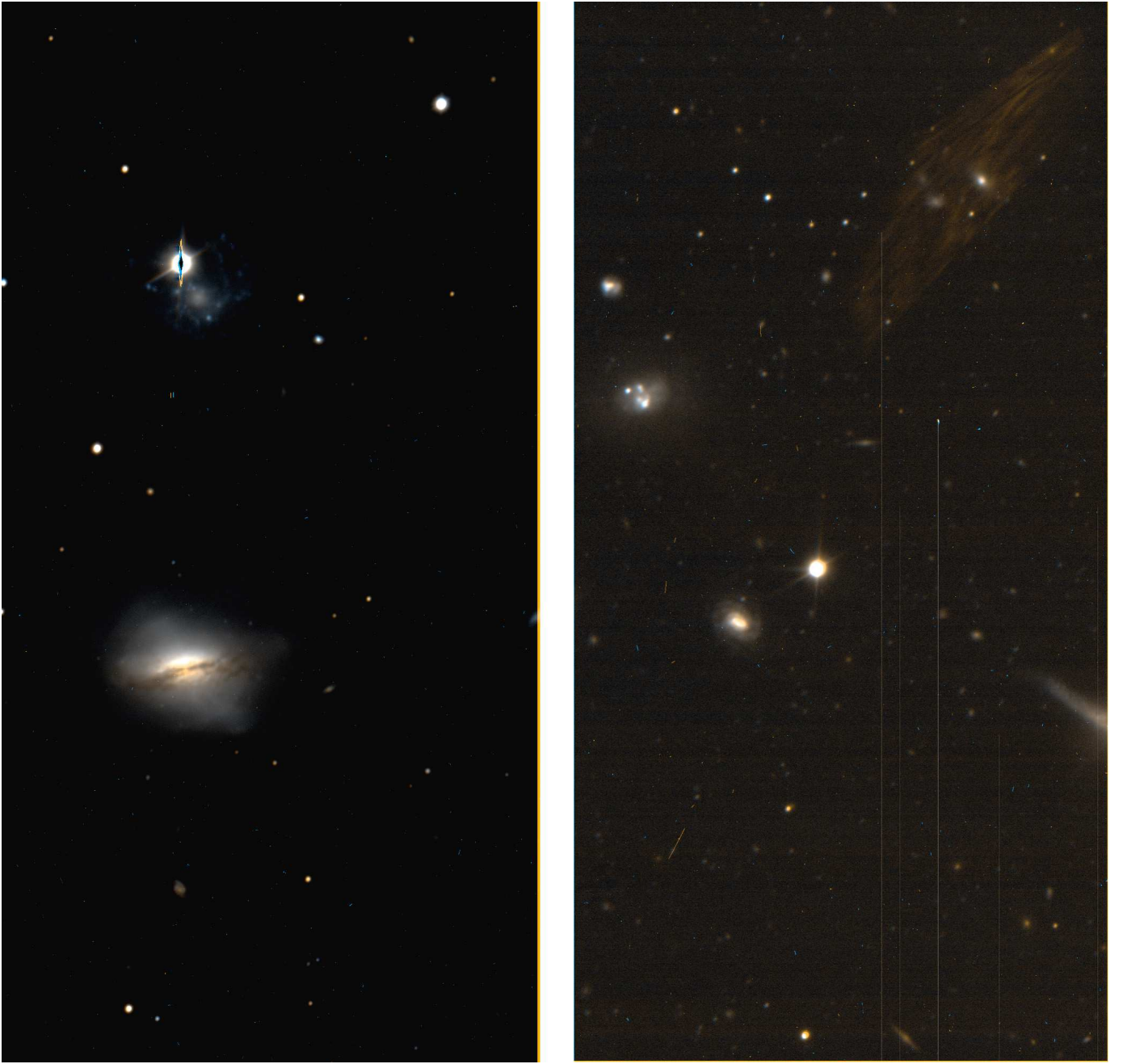}
      \caption{Color composite image of NGC 3934 assembled from TNG+OIG 
      images in the B and R bands, North is up while the East on the left of the 
      image.  In the top right of the figure is visible a defect generated by the CCD
      detector. The field of view is 4.9\arcmin\ $\times$~4.9\arcmin\ . }
               \label{Fig2}
   \end{figure*}
\begin{table*}
\caption{Known members of the NGC~3933 group}             
\label{tabgroup}      
\centering                          
\begin{tabular}{l l ccccc }        
\hline\hline                 
\#  & Name &  RA           &      Decl.   & Morpho.& $V_h$  & B$_T$ \\     
     &           &     [deg]      &   [deg]   & Type      & [km~sec$^{-1}$] & [Vega]\\
\hline                        
1  & NGC 3933 &	  178.00845  &   16.80970    & S0-a &   3734$\pm$10 & 14.27$\pm$0.08 \\   
2  & NGC 3934 &        178.05255  &  16.85163    &Sab  &  3779$\pm$4  & 14.93$\pm$0.27\\  
3  & UGC 6835 &       177.97860   & 16.84582   & --  & 3549$\pm$17 &   16.68$\pm$0.37  \\   
4  &  PGC 213894 &    178.04985  & 16.87857  & &  &  17.78$\pm$0.50\\
5  & SDSSJ115204.21+165217.3 &178.01745  & 16.87164 &--  & 3678$\pm$38 &   17.68$\pm$0.50\\  
6  &  PGC 37112  &  177.90015   & 16.66425  &        --        &    4045$\pm$28 &  16.67$\pm$0.37\\
\hline                                   
\end{tabular}

\noindent{Morphological type,  the systemic heliocentric recession velocity and
the B-band total magnitudes in the Vega-systems are derived from the 
{\it HYPERLEDA} data-base.}
\end{table*}
 
 In this picture, potentially ``evolving" CG systems are interesting because they  
offer the possibility to investigate the multiplicity of mechanisms through which 
member galaxies and the group itself co-evolve. 
Zepf (1993) calculated that roughly  7\% of the galaxies in CGs are
in the process of merging. 

In this context, we here study the properties of the galaxies in the NGC 3933 group.
NGC 3934, the second brighter galaxy in the group,
has been classified as a polar-ring galaxy by Schweizer et al. (1983)
and Whitmore et al. (1990).  Polar-ring galaxies (PRGs) and related objects are 
systems frozen in a peculiar morphology, with matter rotating in two nearly perpendicular planes (see also Bertola et al. 1985 and Varnas et al. 1987).
The {\it Polar Ring Catalogue}  (PRC) lists more than 100 PRGs and 
candidate PRGs (Whitmore at al. 1990). Unfortunately, only a small fraction of 
them has been investigated. There is a general agreement 
that the formation of the polar rings is the result of some
kind of ``secondary event'', such as an external accretion/merger.
Numerical simulations have shown that polar-ring systems could be
explained either by simple gas accretion (Reshetnikov et al. 1997, Bournaud \& Combes 2003) 
or by mergers of perpendicularly oriented disk galaxies (Bekki 1997, 1998).
However, in the PRC most objects do not show this simple 
appearance with an edge-on central object and an edge-on 
polar ring. Instead they appear as systems going toward the stability of a fully mixed and symmetric configuration 
that crosses equilibrium states characterized by irregular structures. These latter systems, called 
in the PRC "polar-ring-related objects", provide a unique chance to investigate
the PRG formation scenarios because the initial components have not 
yet disappeared. If they are the results of mergers, it may  still be 
possible to recognize the progenitors in the unrelaxed remnants, 
and one may attempt to reconstruct their formation  events. 
The NGC~3934 morphology falls in the category of polar-ring-related objects.

On the other hand NGC~3934 also possesses a wide shell system.
It is also widely believed that  shell systems are produced by 
merger episodes  (Dupraz \& Combes 1986, Weil \& Hernquist 1993). 
NGC 3934 is a gas-rich system: it was detected both in HI (van Driel et al. 2002) and  
CO (Galletta et al. 1997) 
and it is a bright IRAS source. It may then represent a case of ``wet'' merging
within compact galaxy associations. This paper presents the properties of
the NGC~3933 system, providing also evidence 
that its ``core" is a compact group of galaxies. 

The paper is organized as follows: photometric and spectroscopic 
observations, data reduction and techniques 
are described in Sect.~2. The results are presented in Sect.~3. 
In Section~4 we discuss the kinematical and dynamical analysis
of the group and interpret the SED and the global properties of the brighter members 
 in the sample, namely NGC~3933 and NGC~3934, 
 using physically motivated SPH simulations. Our conclusions are presented in Section 5.

\begin{table*}
\caption{UV total magnitudes of the group members}             
\label{Table2}      
\centering                          
\begin{tabular}{l c c c c c c }        
\hline\hline   
 Ident.      &   Type  &   major-axis  & minor-axis  &  PA    &   FUV         &  NUV \\ 
                &             &   [arcmin]      &      [arcmin] &  [deg] &    [AB mag]  &[AB mag]  \\     
\hline   
NGC 3933	&   5$\pm$1       & 0.70   &   0.35    &  84.0	  & 17.02$\pm$0.07  & 16.50$\pm$0.04\\
NGC 3934 &  -2$\pm$1       & 0.50  &    0.33   &   70.4  &  20.40$\pm$0.19 & 18.71$\pm$0.06 \\
UGC 6835	&  3$\pm$1        &  0.50   &   0.12     & 64.9   & 19.83$\pm$0.12   & 19.54$\pm$0.08\\
PGC 213894 & 4$\pm$1       &      0.70  &    0.70   &   0.0	  & 17.96$\pm$0.08  & 17.59$\pm$0.04 \\
SDSSJ115204.21+165217.3 &  10$\pm$1 & 0.30   &   0.30   &     0.0 &	  19.01$\pm$0.09 & 18.67$\pm$0.06 \\
PGC 037112 & 8$\pm$1.3 & 0.32 & 0.28 & 2.0 & 18.00$\pm$0.07 & 17.61$\pm$0.04 \\
\hline                                   
\end{tabular}
\end{table*}
 \section{Observations and data reduction}

In Table \ref{tabgroup} we collect the data of the known members of NGC~3933 studied
here.  In the left panel of Figure \ref{Fig1} we show the composite SDSS image of the ``core"
of the NGC~3933 group (galaxies are labeled according to Table \ref{tabgroup}), while in the right panel we
show the composite GALEX image of the same field. PGC 213894, which is faint in the SDSS image is very
bright in the UV one. Contrary to this the presence of dust makes NGC 3934 very faint in the UV image. 

Using the available recession velocities, the NGC~3933 galaxy group 
has  four nearby companions, in projection. Two
of them, NGC 3934 and UGC 6835, have a similar magnitude and size as NGC~3933.
 SDSSJ115204.21+165217.3 and PGC~037112 are noticeably fainter, 
 and the latter  is $\sim$ 10.72$\arcmin$  away from NGC~3933.
At a distance of $\sim 40$\arcmin ~southwest  of the NGC~3933 group "core" 
there is another spiral galaxy,  UGC~6794, with a heliocentric
velocity (3457$\pm$7 km~s$^{-1}$).  {\tt HYPERLEDA} reports
15 additional galaxies within 2 Mpc from the group core and in the recession 
velocity range  3200$\leq V_{hel} \leq$ 4200 km~s$^{-1}$. These are distributed
 along a band oriented east to west; this structure will be discussed in Section~4.1
 
Optical spectroscopic observations have been performed
with the twofold objective to determine the group membership by measuring the recession velocity
of the spiral galaxy PGC~213894, $\sim 1.6$\arcmin\ north of NGC~3934 (with a bright star 
superimposed) and to characterize the velocity dispersion of the polar-ring galaxy NGC~3934. 
\begin{table*}
\caption{Photometric data for NGC 3934, PGC 213894, and SDSSJ115204.21+165217.3. 
We assumed a group distance of 49.62 
Mpc to compute the galaxy absolute magnitudes. Magnitudes are not corrected for galactic extinction.}
\begin{tabular}{lcccccccrr}
\hline
&  &  & & & & & &  \\
   & & \multicolumn{2}{c}{bulge}   & &     \multicolumn{2}{c}{disk}   &  & \\
\cline{3-4} \cline{6-7}
Galaxy & \multicolumn{1}{c}{Band} & \multicolumn{1}{c}{r$_e$} & \multicolumn{1}{c}{$\mu_e$} & & \multicolumn{1}{c}{r$_e$} & \multicolumn{1}{c}{$\mu_e$} &  m$_{tot}$ & M$_{tot}$ \\
name & & [arcsec] & [mag/arcsec$^2$] & & [arcsec] & [mag/arcsec$^2$] & & \\ 
& & & & & & & & \\
\hline
&  & & & & & & & \\
NGC 3934 & B & 9.0 $\pm$ 5.5  & 23.80 $\pm$ 0.88 & & 15.5 $\pm$ 0.2 & 23.50 $\pm$ 0.04  & 14.43$\pm$ 0.29 & -19.29 \\
                & R & 8.0 $\pm$ 4.6  & 21.60 $\pm$ 0.90 & & 16.0 $\pm$ 0.5 & 22.60 $\pm$ 0.10  & 13.03$\pm$ 0.28 & -20.69 \\
PGC 213894 &   B & 22.0 $\pm$ 15.0 & 26.40 $\pm$ 0.78 & & 12.0 $\pm$ 2.1 & 25.00 $\pm$ 0.25 & 15.81$\pm$1.03 & -17.91 \\
                 & R & 22.0 $\pm$ 15.2 & 25.80 $\pm$ 0.80 & & 12.0 $\pm$ 0.4 & 23.80 $\pm$ 0.07 & 14.95$\pm$ 0.34 & -18.77 \\
SDSSJ115204.21+165217.3 &   B & 	- 	& - & & -  & - & 17.30$\pm$0.05 & -16.20 \\
                 & R &       -     &  - & & - & - & 16.59$\pm$0.04 & -16.90 \\
&  &  & & & & & & \\
\hline
\end{tabular}
\label{Table3}
\end{table*}
   \begin{figure*}
   \centering
   {\includegraphics[width=8cm]{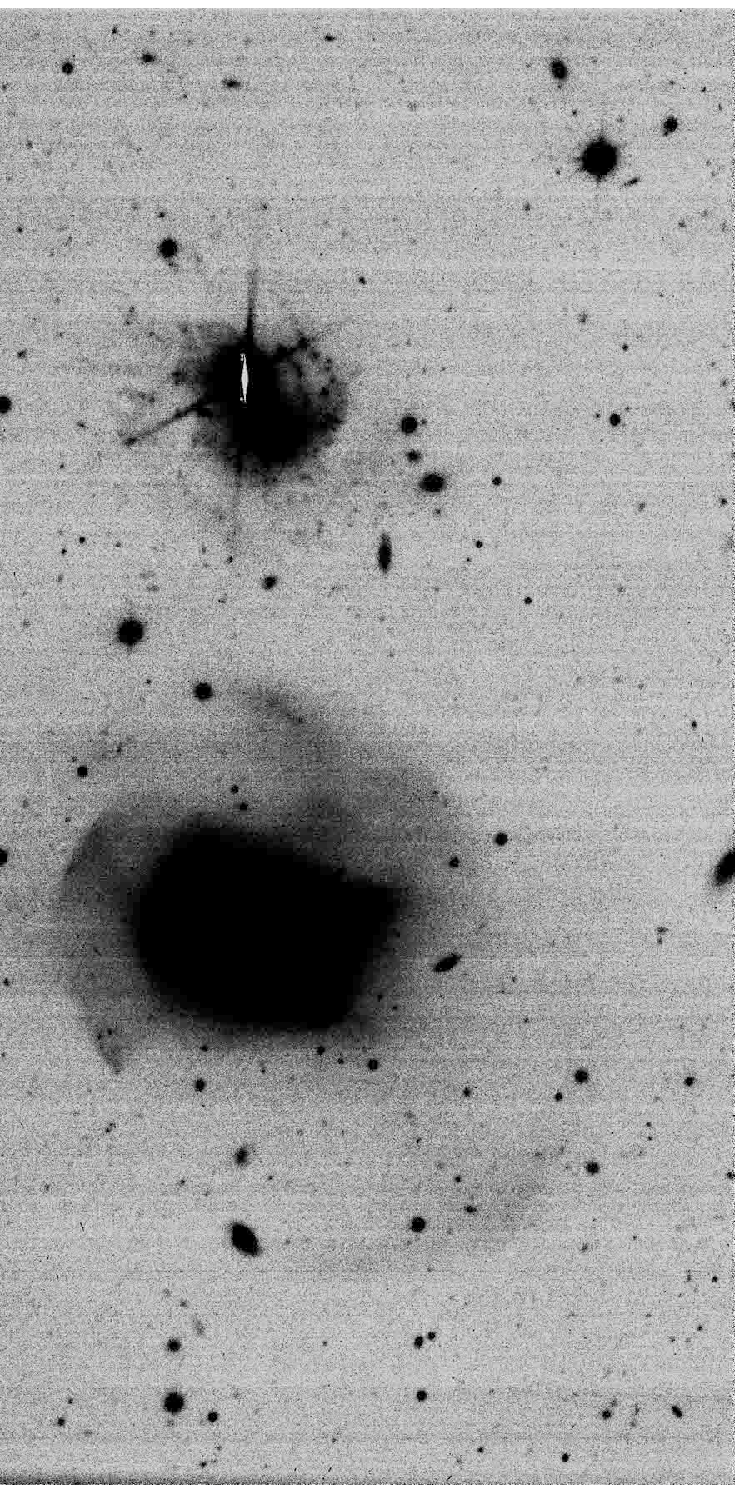}
   \includegraphics[width=8cm]{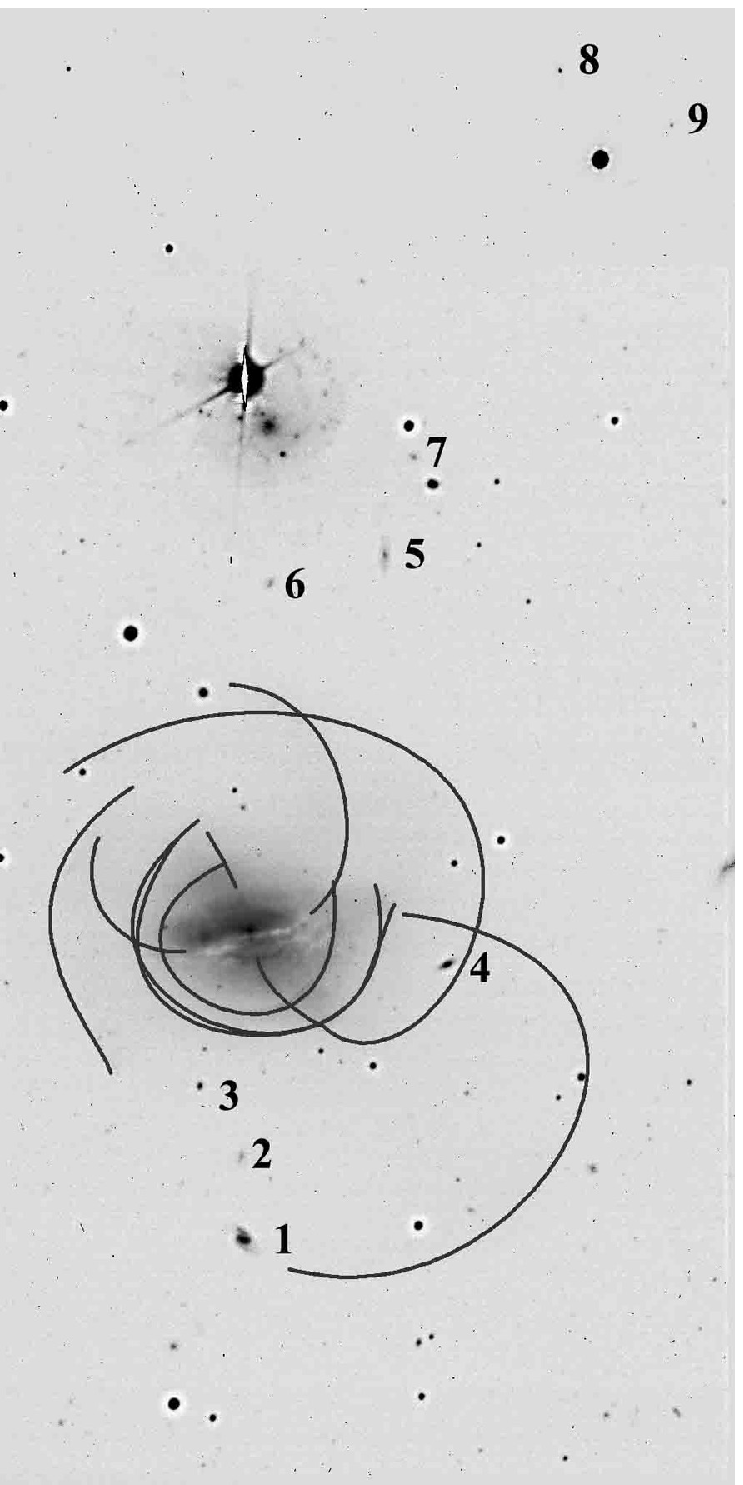}}
      \caption{{\it Left panel}: Deep TNG+OIG R-band image of NGC 3934 
      (North is up, East to the left).  {\it Right panel}: Unsharp masking of
      the R band image. Solid lines sketch the edges of the chaotic system of
      shells and spurs that are visible in the deep image; numbers refers to the faint galaxies listed in Table~\ref{Table3a}.}
         \label{Fig_shell}
   \end{figure*}
   \begin{figure*}
   \centering
   {\includegraphics[width=8cm]{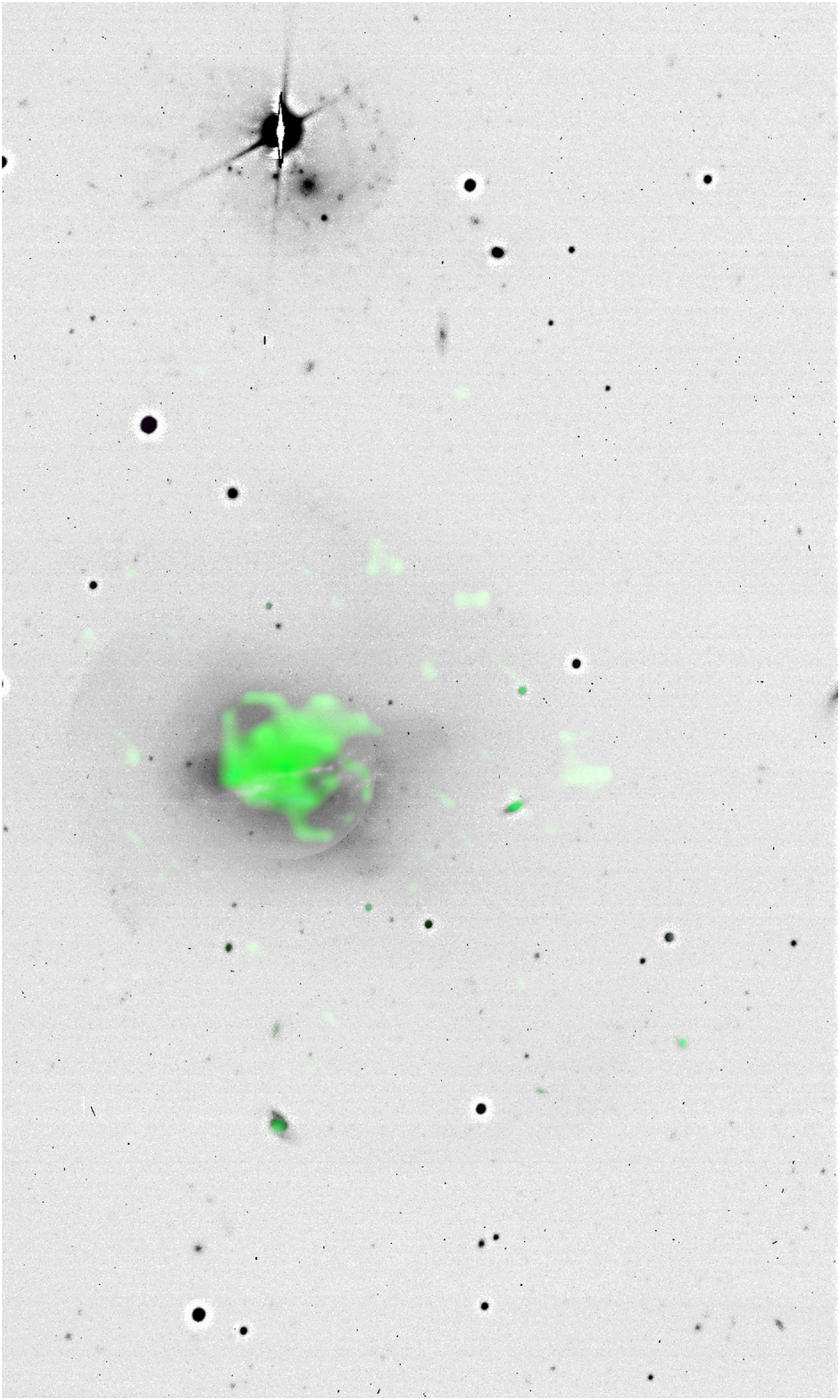}
   \includegraphics[width=8cm]{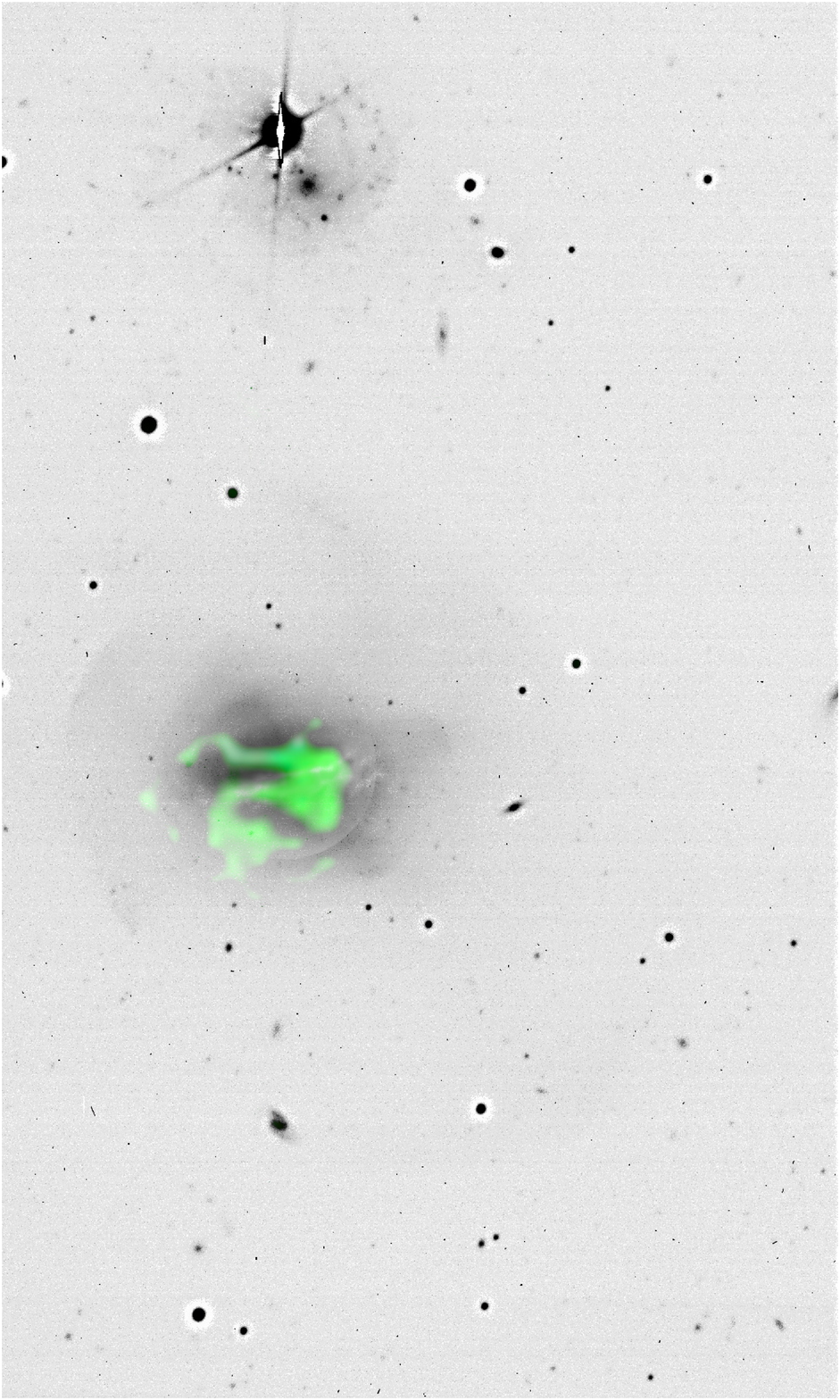}}
      \caption{Unsharp masking of the R-band TNG+OIG image of NGC 3934, 
      superposed the smoothed NUV {\it (left panel)} and FUV  {\it (right panel)} emission
      from {\it GALEX} (see text). North is at the top, East to the left. }
         \label{Fig_UV}
   \end{figure*}

Far UV {\it GALEX} observations and deep optical
imaging were performed to extend  the
galaxy spectral energy distribution to short wavelengths and 
to detail the fine structure of the polar ring of NGC~3934.

\subsection{Spectroscopic observations}

We measured the central velocity dispersion of NGC~3934 
from a galaxy spectrum  taken with ALFOSC@NOT (La Palma, Spain)  
in March 2001. We used ALFOSC which is equipped with a 
Ford-Loral 2024$\times$2024 CCD with 15$\times15\mu$m pixels 
and the grism \#13 with a wavelength range $\lambda\lambda$= 4800-5800\AA. We adopted  a slit width of 2.5\arcsec ~that yields a resolution of 3.3 \AA.The spectrum was obtained along 
PA= 26$^{\circ}$ with an exposure time of 1800s. Two template stars for measuring the radial velocity and velocity dispersion, which are of spectral type G8 III (HD65934) and K0 III (HD132737), were also observed with the same spectral set-up.

In order to measure the redshift of the spiral galaxy PGC~213894,
 we obtained a spectrum at the Asiago Cima Ekar telescope in February 2007. We used 
the Asiago Faint Object Spectrograph and Camera (AFOSC)  
with grism \#4 with a wavelength range $\lambda\lambda$= 3500-7500\AA and a 2.1\arcsec wide slit. This yields a dispersion of 4.9~\AA~pixel$^{-1}$.  The detector is a CCD Tektronix TK 
1024$\times$1024, with pixel size 24$\mu$m, corresponding 
to a spatial scale along the slit of 0.473 arcsec~pixel$^{-1}$. The spectrum was obtained along 
PA= 45$^{\circ}$ with an exposure time of 1800s and also crosses the bright star superimposed on the galaxy. 
The data reduction procedure, performed using the
{\it IRAF}\footnote{{IRAF} is the Image Analysis and Reduction 
Facility made available to the astronomical community by 
the National Optical Astronomy Observatories, which are 
operated by the Association of Universities for Research
in Astronomy (AURA), Inc., under contract with the U.S. 
National Science Foundation.} {\it LONGSLIT} package, was 
similar for both spectra. Scientific frames were corrected for bias and flat-field
and calibrated using an He/Ar arc-lamp.

No emission lines are present in the spectrum of NGC 3934 
which was very noisy, because of  dust in the central regions. For this reason, 
we co-added the spectrum in an aperture of 10\arcsec.  This may lead to an overestimate of the measured velocity dispersion if a rotation gradient is present. With our adopted distance, the coadded region corresponds to a physical region of 2.5 Kpc; for an early-type galaxy this can lead to an overestimate of about $\sim10$\%, well within the errors of our measure. The velocity dispersion and the radial velocity were derived, 
using the Fourier quotient technique (Sargent et al. 1977, Bertola et al. 1984). We obtained the following results: 
$V_r$=3740$\pm$88 km~sec$^{-1}$ and $\sigma$=169.5 
$\pm$19.5 km~sec$^{-1}$. The galaxy recession velocity, corrected for heliocentric motion,
agrees with values already available in the literature
(see Table \ref{tabgroup}).  

The spectrum of the spiral PGC~213894 shows  the $O[III] \lambda$5007,
$H_{\alpha}$ and [NII] $\lambda\lambda$6548-6583 emission lines.  All emission lines were
fitted with a Gaussian profile and the final measured galaxy radial velocity is the mean value with the error being the standard deviation. The velocity, corrected to the Sun, is $V_r$=3542$\pm40$ km~sec$^{-1}$, which is
compatible with the average group velocity (see Table \ref{Table4}) and implies that the galaxy is a member of the group.

\subsection{UV and optical imaging}

The UV imaging was obtained with GALEX observatory 
(see Martin et al. 2005; Morrissey et al. 2005) 
in its ultraviolet bands FUV (1344 -- 1786 \AA) ~ and NUV (1771 -- 2831
\AA). The instrument consists of a 50 cm diameter modified Richey-Chre\`tien telescope
that has a very wide field of view (1.25 degrees diameter) and a spatial 
resolution of $\approx$4\farcs2  and 5\farcs3 FWHM in FUV and NUV
 respectively, sampled with 1\farcs 5$\times$1\farcs 5 pixels.  
 
The FUV and NUV images were obtained on March 27, 2006, by 
means of dedicated  observations awarded to our team (GI1 program 59, P.I. D. Bettoni). 
The exposure times were 1683 sec in both bands
 (limiting magnitude in FUV/NUV $\sim$ 22.6/22.7 AB mag (Bianchi 2009). A UV composite image of the group (FUV blu; NUV yellow) is shown in the right panel of Figure \ref{Fig1}. 

We used FUV and NUV {\it GALEX} background-subtracted intensity 
images to compute the integrated photometry of the galaxies within elliptical apertures
(see Table~\ref{Table2}). Background counts were estimated from the sky-background image and  the 
high-resolution relative response map 
provided by the {\it GALEX} pipeline. 
In Table \ref{Table2} we report the morphological type following de Vaucouleurs et al. (1991), the major and minor axis length, the position angle of the major axis, and the total FUV and NUV magnitudes of the group members. They were computed as m(AB)$_{UV}$ = -2.5 $\times$log CR$_{UV}$ + ZP, where CR is the dead-time corrected, flatfielded count rate, and the zero points are ZP=18.82 and ZP=20.08 in FUV and NUV, respectively (Morrissey et al. 2007). 

   \begin{figure}
   \centering
   \includegraphics[width=7.5cm]{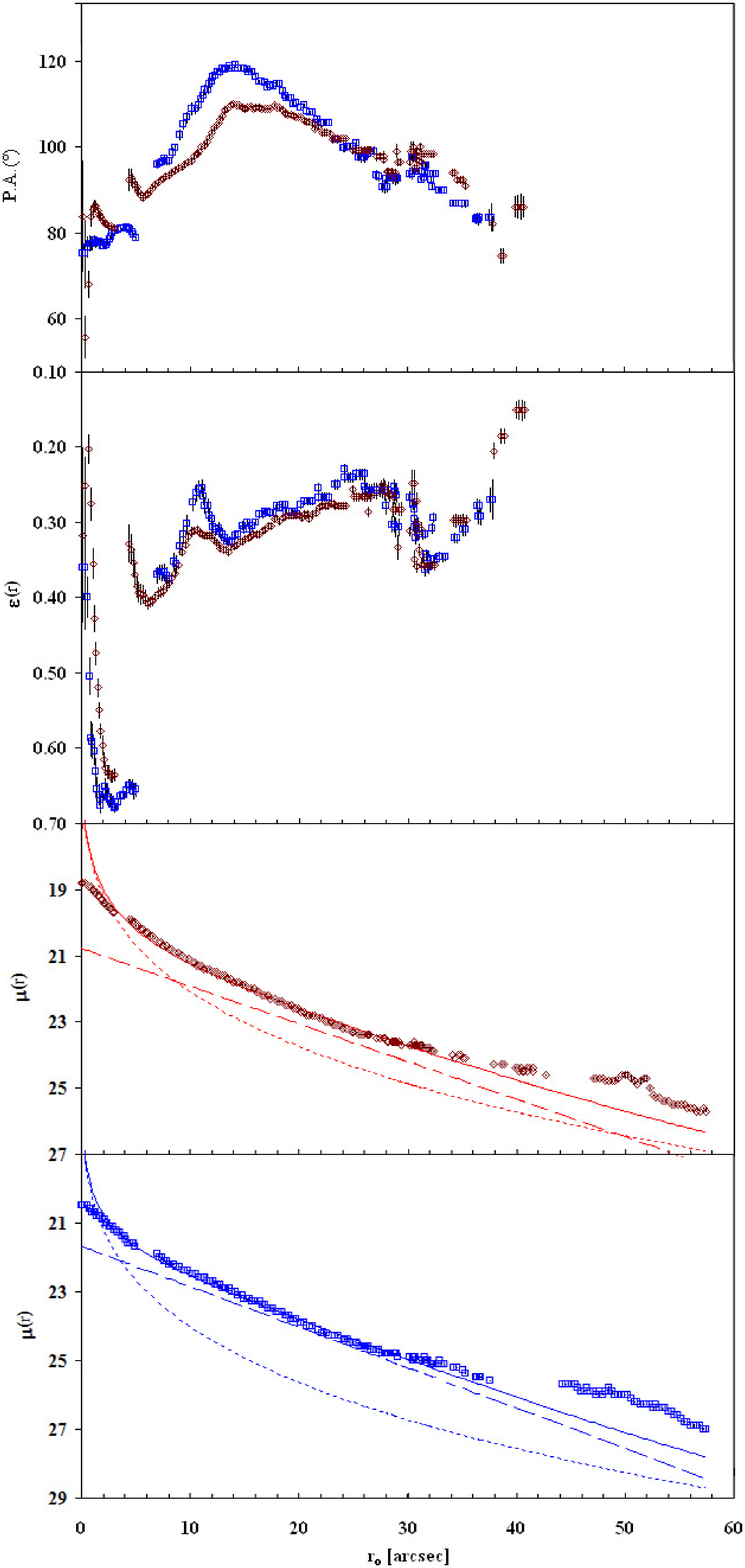}
      \caption{NGC~3934: position angle (top panel), ellipticity (mid panel), 
      and surface brightness (bottom panels) profiles as function of the
      galacto-centric distance in the B- (blue open squares) and R- 
      (red open diamonds) bands obtained from the TNG+OIG observations. In addition to
      the global luminosity profile, the profiles of disk (dashed line) and bulge (dotted line) are also shown.}
         \label{Fig_prof}
   \end{figure}
   \begin{figure}
   \centering
   \includegraphics[width=8.5cm]{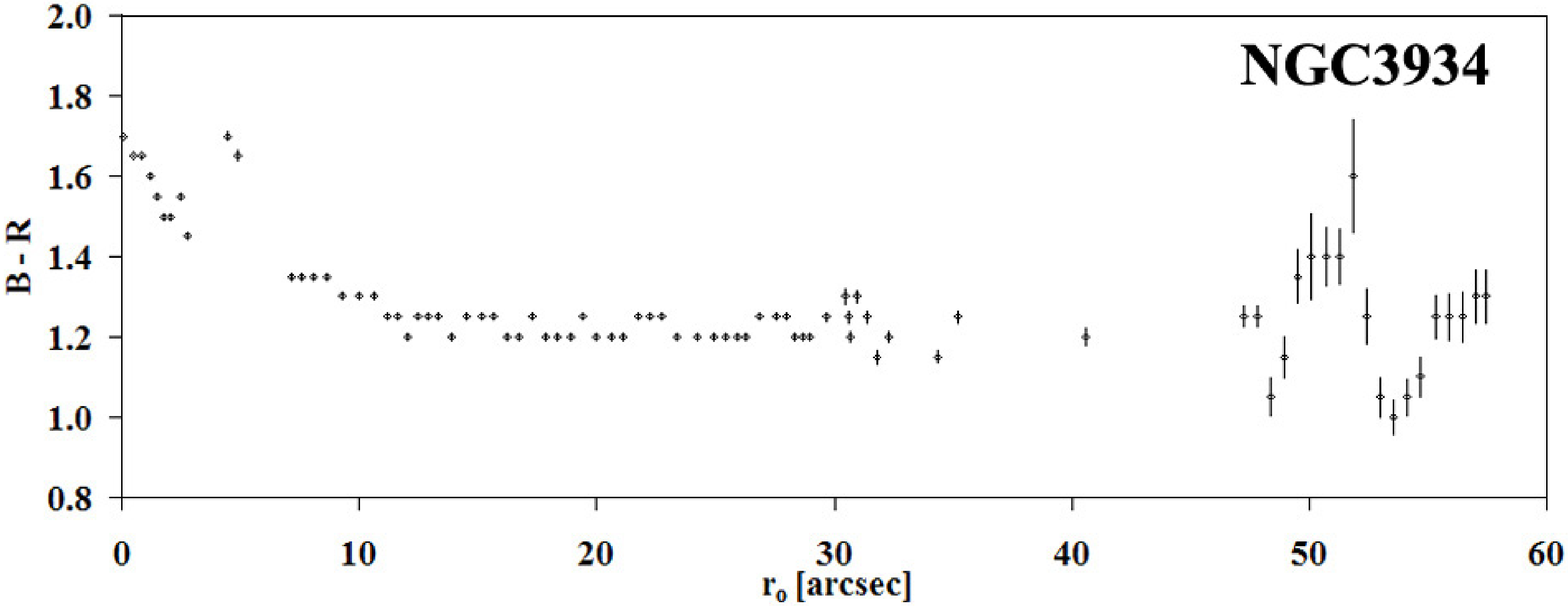}
      \includegraphics[width=8.5cm]{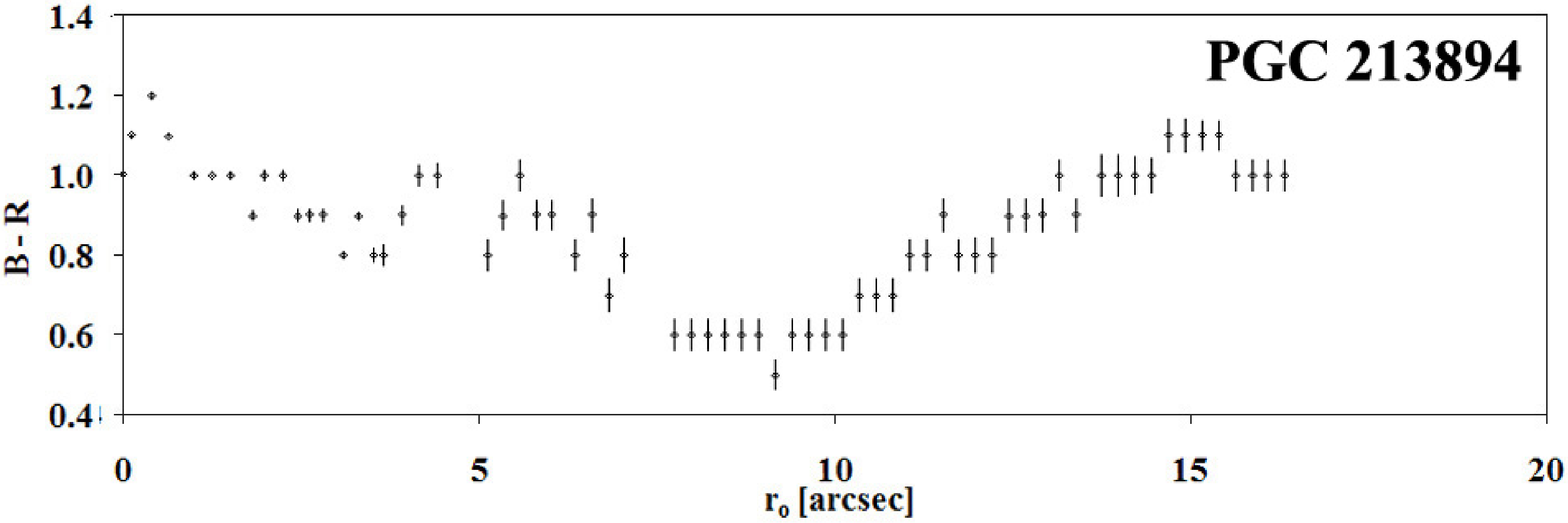}
      \caption{Top panel: color profile of NGC 3934; bottom panel: color profile of  PGC 213894.}
         \label{Fig_col}
   \end{figure}
 
Optical photometric observations were carried out in May 2000 with
OIG@Telescopio Nazionale Galileo. The OIG was equipped
with a mosaic of two thinned and back-illuminated EEV 42-80 CCDs,
each with $2048 \times 4096$ pixels. The pixel size and pixel
scale are 13.5$\mu$m and 0.072 \arcsec/pixel
respectively, corresponding to a total field of view of about 
4.9\arcmin$\times$4.9\arcmin. The OIG data were resampled with a 
binning factor of two, corresponding to a spatial scale of 0.144\arcsec/pixel.
Exposure times were 1200s in the R- and B-Johnson band filters.
Photometric calibration was achieved using the
observations of standard stars from the Landolt (1983) and Smith et al. (1991) lists. The seeing during the observations was $\sim$1.2 arcsecs. Figure \ref{Fig2} shows the color image of the OIG frame.

The standard CCD image reduction was performed  using the 
{\it IRAF} package. This includes dark and bias subtraction, flat-field correction
(we used a mean of several dome flats taken in the appropriate
filter), and sky subtraction. The cosmetic defects and cosmic
rays were excluded by median filtering and the projected stars
by masking them with rectangular regions. The typical
uncertainty of the background level is about 0.5\%.
   
The surface photometry was derived by means of the {\it ELLIPSE} 
package (Jedrzejewski 1987), which computes a Fourier expansion for each
 successive isophote, producing radial luminosity and 
 geometrical profiles sampled in nested ellipses. These profiles are plotted 
 versus galacto-centric distance along the circularized semiaxis in Fig. \ref{Fig_prof}.
 

 \begin{table}
\caption{Magnitudes and colors of the faint galaxies in the field}             
\label{Table3a}      
\centering                          
\begin{tabular}{l c c c }        
\hline\hline   
 Ident.      &   $m_R$  &  $m_B$  & B-R  \\ 
\hline     
1  & 19.32$\pm$0.073  &    21.17$\pm$0.07 &   1.78$\pm$0.101  \\
2  & 21.41$\pm$0.212  &    22.62$\pm$0.08 &   1.21$\pm$0.226  \\
3  & 21.27$\pm$0.042  &    22.95$\pm$0.14 &   1.68$\pm$0.146  \\
4  & 20.24$\pm$0.103  &    22.02$\pm$0.12 &   1.78$\pm$0.158  \\
5  & 20.91$\pm$0.030  &    22.19$\pm$0.09 &   1.28$\pm$0.094  \\
6  & 21.45$\pm$0.100  &    23.16$\pm$0.13 &   1.71$\pm$0.164  \\
7  & 21.60$\pm$0.049  &    23.31$\pm$0.08 &   1.71$\pm$0.094  \\
8  & 20.02$\pm$0.055  &    22.28$\pm$0.20 &   2.26$\pm$0.207  \\
9  & 20.40$\pm$0.076  &    22.66$\pm$0.13 &   2.26$\pm$0.150  \\ 
\hline                                   
\end{tabular}
\end{table}

We obtained the detailed surface photometry of both  NGC~3934
and of the new member of this group, PGC~213894. In order to extract 
the  effective radius r$_e$ and the effective surface brightness 
$\mu_e$ of these galaxies, we performed a best fit of their surface brightness profiles.
The images, the luminosity, ellipticity, and position angle profiles suggest that
these galaxies have a disk component in addition to the bulge. We then    
use a model composed of a bulge, following the r$^1/4$ law 
(de Vaucouleurs 1948), and an exponential disk (Freeman 1970). 

PGC~213894 has a bright star (SDSS J115212.60+165255.4) 
at 11\arcsec\ NE of the nucleus. This star, showing the typical spectral 
features of an M star, makes it difficult to fit ellipses to the galaxy image. 
In addition, the disk appears faint and the spiral arms dominate the B- and 
R- images. To overcome these problems, we fitted not only the circularized 
luminosity profiles, but also the aperture photometry 
produced by {\it ELLIPSE}. We made on our theoretical 
profiles fit both the luminosity profile, $\mu(r)$,
and the growth curve, m(r). The central absorption is then minimized 
because it contributes less to the total value of integrated magnitude. 
Results of the best fitting, reported in Table \ref{Table3}, agree
with those available in {\tt HYPERLEDA}\footnote{http://leda.univ.lyon1.fr/}. 
In the right part of the OIG frame, 1.1\arcmin ~SW of SDSSJ115204.21+165217.3, the galaxy SDSSJ115202.45+165117.4 is also visible, a background object with a redshift z=0.115. Moreover many faint galaxies are visible in our image that may be related to the group. 
We measured the total B- and R- magnitudes of these faint object and report them in Table~\ref{Table3a}. 
Despite their late-type morphology their (B-R) color appears to be much redder than that of their morphological class (see e.g. the color profile of PGC 213894 in Figure~\ref{Fig_col}). We suggest that they are background galaxies. 
 
\section{Results}

\subsection{Morphology and structure of member galaxies from optical OIG observations.}

NGC~3934 has strong dust--lanes that irregularly obscur the inner and brighter regions
of the galaxy. They are shown in Figure \ref{Fig_shell} (right panel). Spurs and shells are
visible in the inner parts, but especially in the outskirts of the galaxy. The B-R color profiles for NGC 3934 and PGC 213894 are shown in Fig.~\ref{Fig_col}. In NGC 3934, the galaxy outskirts have an average B-R$\sim$1.2, that is not as blue as typical spiral arms although they are bluer than the central parts which is dominated by the dust--lane, . As an example in the nearby spiral companion PGC 213894, where the average color of the arm region is $\sim$0.9.

This irregularity is also visible from the data shown in the top panel of Figure \ref{Fig_prof} where  there 
is a rapid increase of the position angle ($\Delta$P.A.$\sim$40$^{\circ}$) in the inner 15\arcsec\ , followed by a 
monotonic decrease in the outskirts. Summarizing,
we do not find signatures of either a polar-ring structure or of an inclined ring
in the galaxy; the system is instead embedded in a complicated, type-II  systems 
of inner and outer shells (see Prieur 1990). This shell 
system  is reminiscent of that observed in the interacting
 S0 galaxy NGC~474  (see Figure~1 in Rampazzo et al. 2006).
 However, the strong dust--lane in addition
 to the system of shells  points out  that NGC 3934 is the product of  a ``wet'' merging episode.

The B- and R-band luminosity profiles of NGC~3934  show a disk. In the B-band we measure a typical disk central brightness $\mu_0=21.68$, 
similar to the classic value of 21.65$\pm$0.30 found for a disk (Freeman 1970) with a scale length of 9.2\arcsec (see data in Table \ref{Table3}). Our total B- magnitude m$_B$=14.43$\pm$0.29 agrees with the values  
m$_B$=14.27$\pm$0.08 listed in the {\it HYPERLEDA} catalogue. 
 The total absolute  magnitude indicates an intermediate luminosity galaxy 
 (see e.g. Sandage \& Tamman 1987).
 
PGC~213894, discovered to be associated to the NGC~3933 group, is an
intermediate luminosity Sbc spiral with a faint disk. We estimated a total B-
magnitude m$_B$=15.81$\pm$1.03 quite different from the value given by {\it HYPERLEDA}. 
The large error on the B- magnitude is attributable
to the difficulty of the fit created by the residuals of the foreground M star, even after its subtraction.
 No obvious signatures of morphological distortion are visible in the disk of the galaxy.

The high-resolution OIG images permit us also to better describe the
morphology of SDSSJ115204.21+165217.3. This faint galaxy has an
irregular appearance with bright knots irregularly distributed within a faint
envelope. Because of this clumpy structure we did not attempted to fit any
luminosity profile; to establish its total magnitudes, we preferred to fit 
the growing curve of the magnitude within a fixed radius (m$\leq$ r),
extrapolated to infinity. Indeed,  we report the total magnitudes only in Table \ref{Table3}.
 The comparison with the stellar position in GSC3 
catalogues shows that knots are likely part of the galaxy, because these are extended sources.   

The edge-on spiral UGC~6835 shows a warped disk structure. 
This feature and the NGC~3934 system of shells are the only morphological unambiguous 
signatures of interaction  in the NGC~3933 group of galaxies.

   \begin{figure}
   \centering
  \includegraphics[width=7cm, angle=-90]{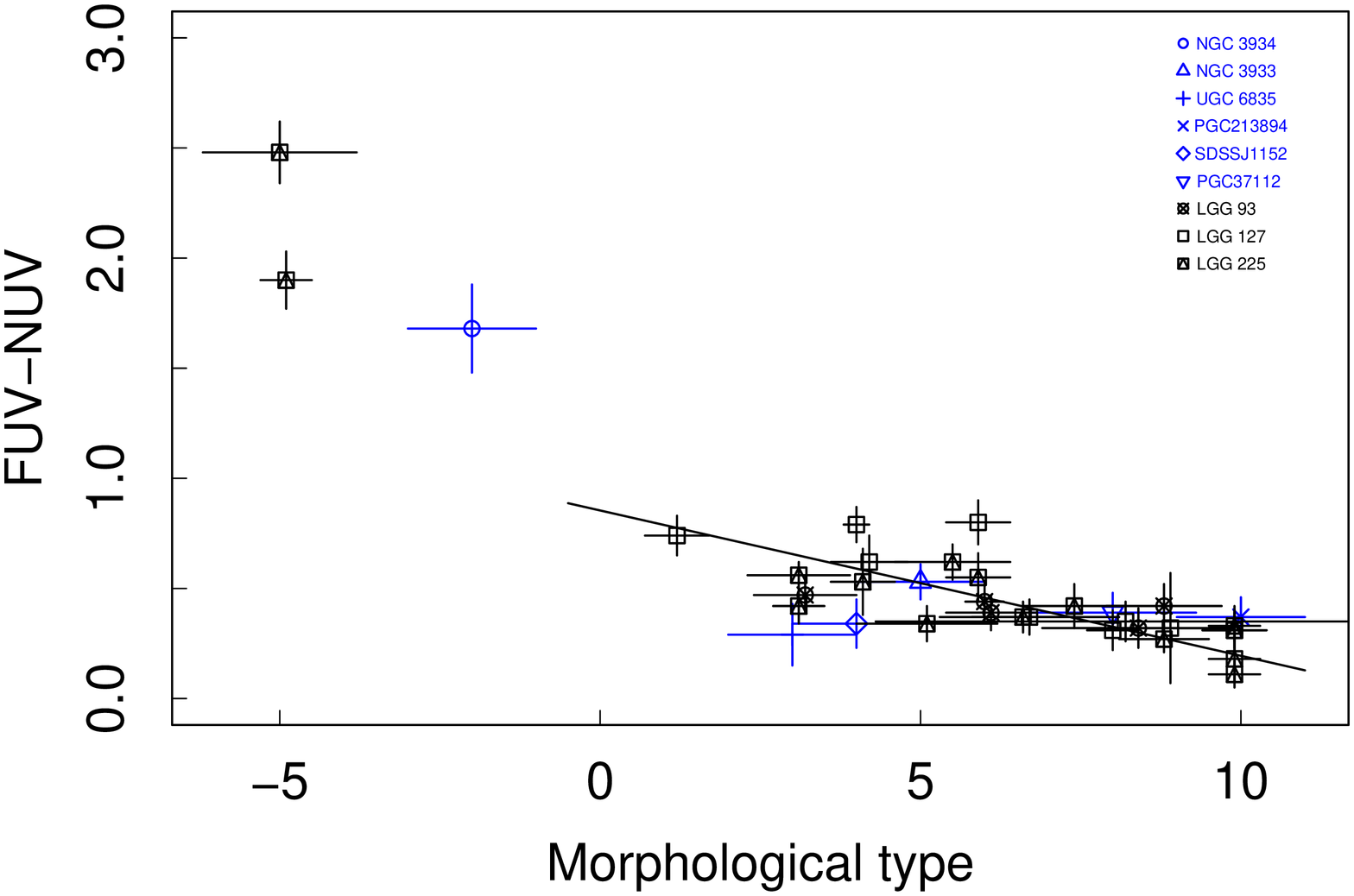}
  \includegraphics[width=9cm, angle=-90]{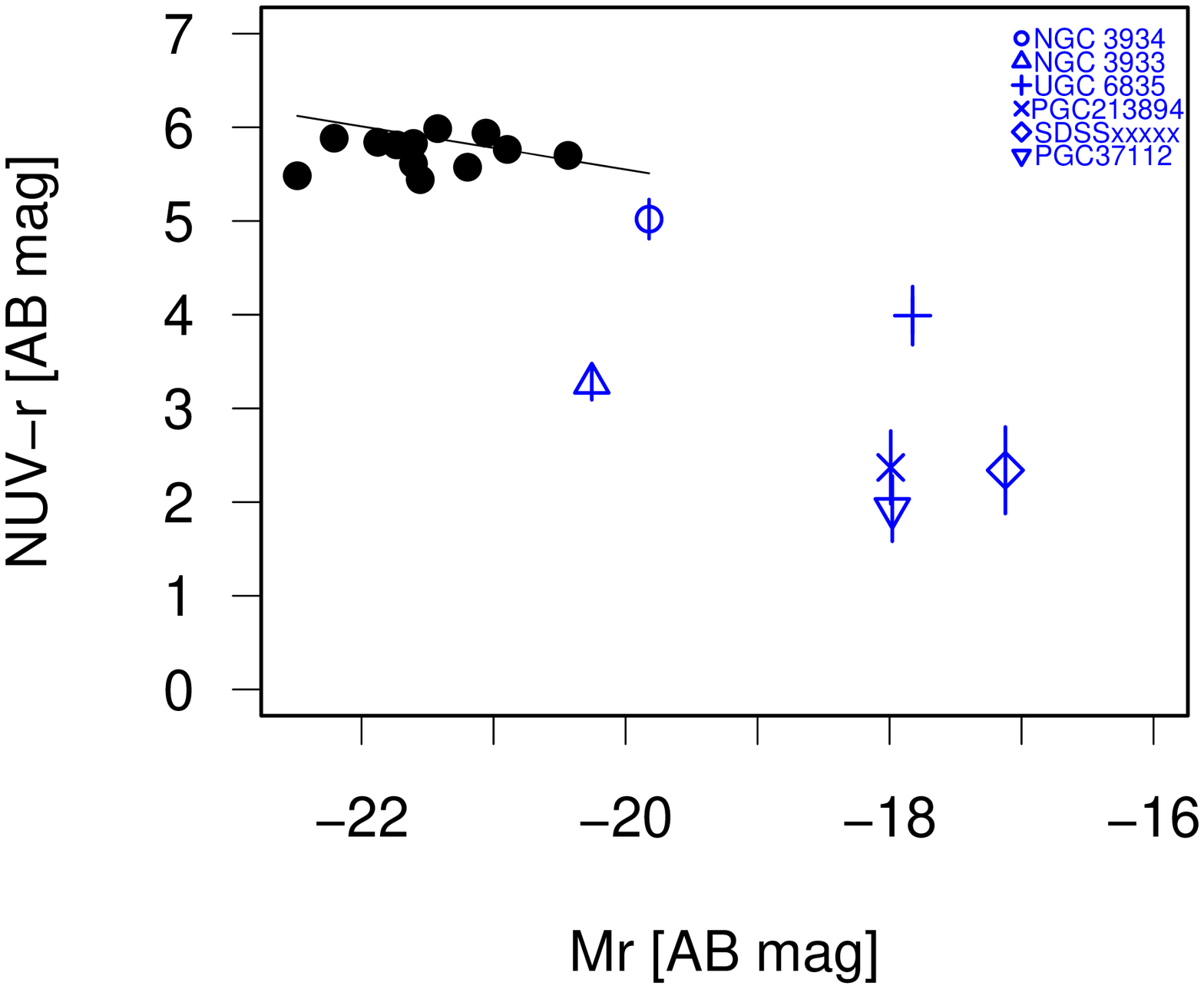}
      \caption{{\it Top}: (FUV-NUV) color as a function of the morphological type for
      the six group members of the NGC~3933 group. The solid line represents the
      best linear fit for types $T=-0.5$ or later (i.e. spiral galaxies and irregular) in
      Gil de Paz et al. (2007). Members of LGA groups from Marino et al. (2010) are plotted for comparison (see text). {\it Bottom}: NUV-$r$ vs M$r$
      color-magnitude plot for galaxies in the NGC~3933 group. The Yi et al. 2005)
      red sequence is shown as a solid line. Early-type galaxies in Marino et al. (2011)      are indicated as black dots (see text).  }
         \label{Fig_env}
   \end{figure}

\subsection{UV morphology and integrated properties}

In Figure \ref{Fig_UV} we plot the NUV and FUV emission over the 
unsharp masking of the R-band OIG image. We notice that 
the FUV emission is centered on the nuclear part of the
galaxy and the dust-lane strongly absorbs the emission
in this area. The NUV emission is still in the nuclear part of the
galaxy but more homogeneously distributed and extended. 
Furthermore, we notice some NUV knots in the northwest side of the 
galaxy along the edge of the shell. These NUV knots along the NW ripple of NGC 3934 
offer a similarity with another shell galaxy, NGC 1210 (Marino et al. 2009). In this latter galaxy several knots, visible both in the NUV and the FUV bands, follow a tail-like structure in coincidence with a HI tail/ring. 
This point  further supports our interpretation of these features in NGC 3934 as the 
debris of a recent accretion event that is responsible for the shell structure.

In Table \ref{Table2} we present the values of the integrated photometry 
in the FUV and NUV bands of {\it GALEX}. In the second column
we also report the morphological classification we adopted for the
galaxies in the group. In general, the morphology is not provided in the
current literature (see Table \ref{tabgroup}) or is incorrect, as for NGC~3934,
which we reclassified as a peculiar S0 on the basis of the present 
surface photometry. 

In Figure \ref{Fig_env} (top panel) we plot the (FUV - NUV) color as a function of the 
morphological type.  The average relationship found
by Gil de Paz et al. (2007) for late--type galaxies (from spirals 
to irregulars) is over-plotted as a solid line. For comparison
we also plot the members of Local Group Analogs (LGA) analyzed
by Marino et al. (2010). These groups are mostly composed
of late-type members, although two early-type are present.  
The galaxies in the present sample are distributed in this plane 
with a dispersion consistent with the findings of
Gil de Paz et al. (2007) and with colors similar to groups
studied by Marino et al. (2010).  The (FUV-NUV) color of NGC 3934, 
greatly absorbed by strong dust-lanes, locates
 the galaxy  in the region nearby early-type galaxies in LGA group.
 
 In Figure \ref{Fig_env} (bottom panel) we show the NUV-$SDSS-r$ color vs. the absolute M$r$
 magnitude. For comparison we also plot the set of early-type
 galaxies in Marino et al. (2011), for which both {\it GALEX} and
 SDSS observations were available. These latter are located on the
 red sequence derived by  Yi et al. (2005)
  [(-0.23$\pm$)0.30$\times$ M$_r$ + 0.75] for early-type
 galaxies in the local universe (0$< z <$0.05). Late-type galaxies, in particular NGC 3933 and UGC 6835, are located in the  blue sequence but also in the green-valley (Salim et al. 2007).
 The shell galaxy NGC 3934 with its large absorption bands
  is also located in the green valley.  Note that galaxies showing morphological signatures of interaction
  are located in the transitional region, the green valley, where evolving systems are found (Marino et al. 2011a).

   \begin{figure*}
   \centering
   \includegraphics[width=\textwidth]{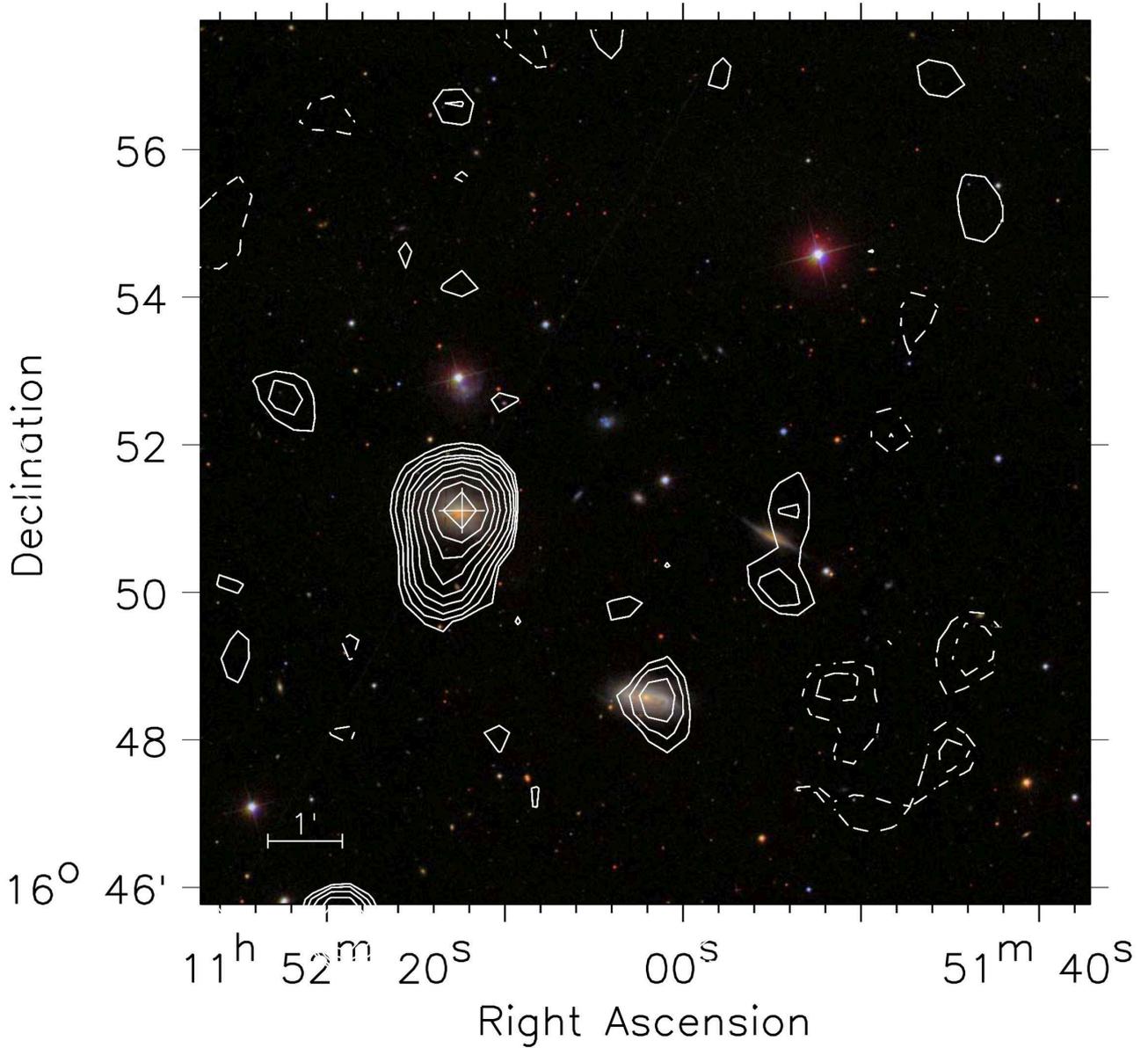}
      \caption{Composite image from the SDSS survey of the field of the 
      NGC~3933 group with the NVSS  map  superimposed. (Levels = 1, 1.4, 2, 2.8, 4 ... mJy/b)}
         \label{Fig_radio}
   \end{figure*}


 In Figure \ref{Fig_radio} we show the SDSS composite image of the NGC~3933 group
with the isocontours of the 1.4 GHz continuum emission from the
NRAO VLA Sky Survey (NVSS, Condon et al. 1998) superimposed. 
Two galaxies in the field are associated with significant radio emission. From Condon et al. (1998) we derive for NGC 3934 F$_{1.4}$=2.8$\times10^{-28}$ W$m^{-2}Hz^{-1}$ corresponding with our adopted distance (see Table \ref{Table4}) to a luminosity L$_{1.4}$=8.34$\times10^{21}$ W$Hz^{-1}$ and for NGC 3933 F$_{1.4}$=3.0$\times10^{-29}$ W$m^{-2}Hz^{-1}$, corresponding to a luminosity L$_{1.4}$=8.94$\times10^{20}$ W$Hz^{-1}$. The ratio L$_{1.4}$(NGC3934)/L$_{1.4}$(NGC 3933)$\sim$10 is the same as the one we found for the 60$\mu$m flux ratio. The 1.4GHz luminosity is insensitive to dust obscuration and for this reason is a good tracer of the star formation rate (SFR$_{1.4}$). We adopt the calibration of Hopkins et al. (2003) and we found an SFR=4.6~M$_\odot$~yr$^{-1}$ for NGC 3934 and SFR=0.83~M$_\odot$~yr$^{-1}$ for NGC 3933. These SFR will be discussed below.

   \begin{figure*}
   \centering
   \includegraphics[width=18cm]{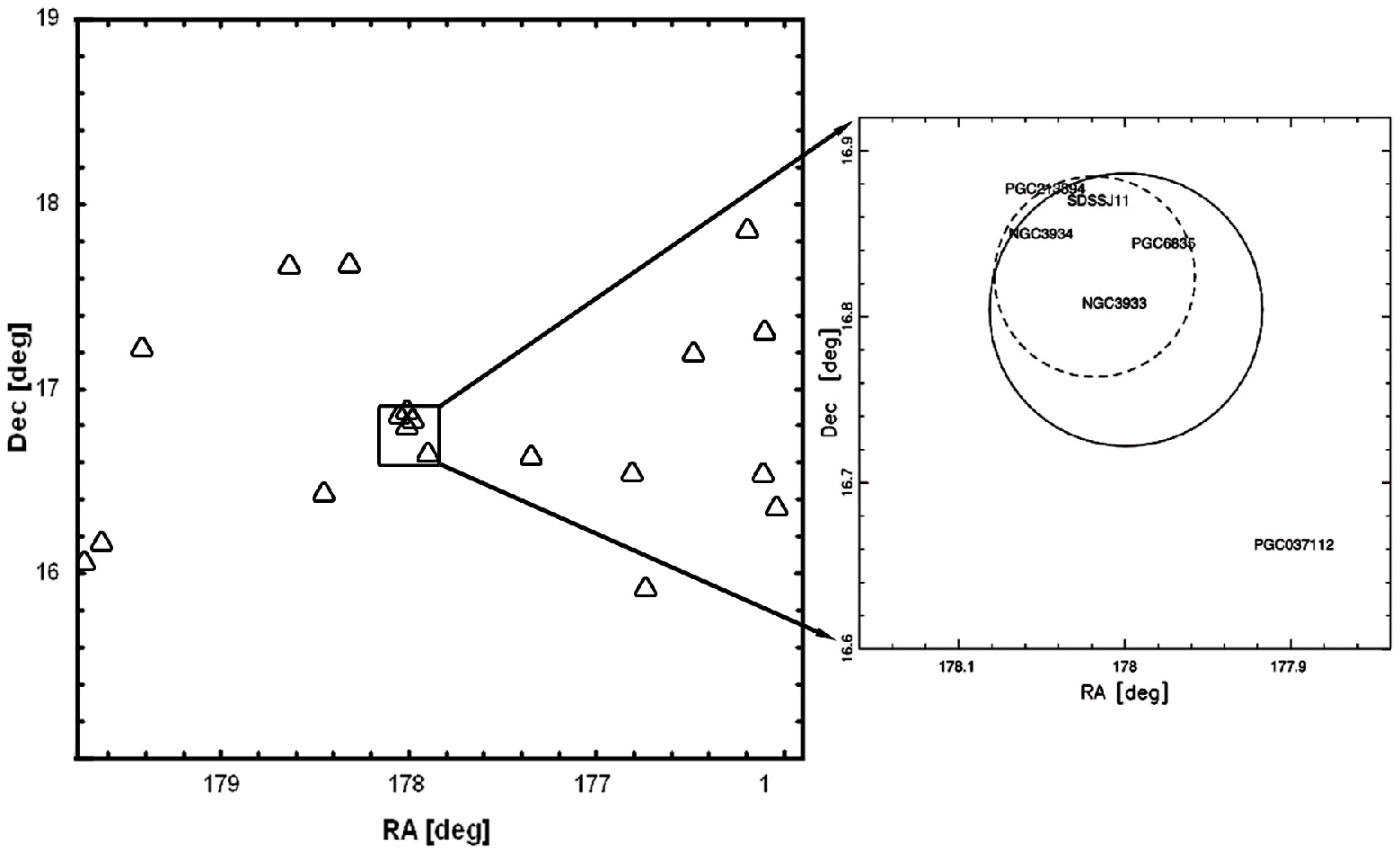}
      \caption{Left panel: Distribution of the galaxies within 2 Mpc and the recession velocity
      range 3200 $\leq V_{hel} \leq$ 4200 km~s$^{-1}$. Right panel: Zoom on the NGC 3933 group
      (FoV 19\arcmin$\times$19\arcmin). 
      The solid and dashed lines show the harmonic radius in the NUV and $r-$ bands,
      respectively (see Table \ref{Table4}).}
         \label{Fig_group}
   \end{figure*}

\section{Discussion}

\subsection{Group membership, kinematic and dynamic analysis 
of the NGC~3933 group} 

The projected position and the luminosity properties
of five out six galaxies, (Table \ref{Table4})  fulfill the criteria devised by 
Hickson (1982) for
a compact group of galaxies.  We indicate these galaxies
as the ``core" of the NGC 3933 group, as shown in Figure~\ref{Fig_group}, where we plot the 
projected distribution of galaxies within 2 Mpc and $\pm$500 km~s$^{-1}$
from the average recession velocity of the group itself. The galaxies
included in the larger box in Figure~\ref{Fig_group} are PGC 139672, PGC 036501, PGC 213877, PGC 036524,
NGC~3853, PGC 036561,  PGC 036681, PGC 1496906, PGC 036768, UGC 06794, PGC 1538830,
SDSSJ115348.42+162636.0, PGC 2806925, SDSSJ115744.10+171329.1, NGC 4014, and
SDSSJ115857.18+160404.0. According to the {\tt HYPERLEDA} classification,
these galaxies are mainly spirals, like in the NGC 3933 group core.
Ramella et al. (1994)
suggest that compact configurations are continuously forming 
within loose groups. 


Our multi-band photometric data allow us to perform  
a multi-band kinematical and dynamical analysis of the group.
Table \ref{Table4} summarizes our results computed according for the
recipes given by Firth et al. (2006). All mass-related quantities are obtained by
luminosity-weigthing the contribution from each member. Each galaxy is weighted by its relative luminosity
evaluated from FUV to r magnitude converted to relative luminosity  
(with the following absolute solar magnitudes: M$_{FUV\odot}$=16.02, 
M$_{NUV\odot}$=10.15, M$_{B\odot}$=5.45, M$_{r\odot}$=4.67\footnote{http://mips.as.arizona.edu/~cnaw/sun.html}). The average center of mass, weighted on luminosity, 
shifts by about 1.5\arcmin\ from the optical to UV bands, moving 
southwest from  NGC~3933 towards PGC~37112.
The right panel of Figure~\ref{Fig_group} shows the galaxies included in the harmonic radii of NGC 3933, weighted at NUV and $r-$band magnitudes.

The  projected mass  increases from 2.5$\pm$0.9$\times$10$^{12}$
M$_\odot$ in the optical bands to 7.7$\pm$2.5 $\times$10$^{12}$ M$_\odot$ 
in the FUV.  A  crossing time of 0.04 Hubble 
times suggests that, at least in the central region, the group 
is virialized.  The virial mass of the group is small, lower than 
2.3$\pm$0.9 $\times10^{12}$ $M_\odot$. Using this mass value,
the relationship by Mathews et al. (2006) predicts a diffuse X-ray luminosity of 
about 10$^{39}$\,ergs~s$^{-1}$ compatible with upper limits of the Rosat All Sky Survey (RASS, Voges et al. 1999). The overall group velocity 
dispersion on the order of 100-150 km~s$^{-1}$ is  relatively low,  but not 
unusual for compact configurations (see e.g. Mamon 2000). This low
velocity dispersion could either argue against the virialization of the group
(chance alignment with a loose group) or suggest that the tidal friction 
is slowing down the galaxies and bring the group toward the full coalescence.   

\begin{table*}
\caption{Dynamical analysis of the NGC 3933 group}
\label{Table4}
\centering
\begin{tabular}{l c l l l l l l l l}
\hline\hline
band    &     Center       & V$_{group}$      &   Group      &Dist.$^+$  &  Harmonic   & Virial & Projected & Crossing  & Group \\
             &         of mass  &                            & vel.  disp.     &         & radius         &   mass & mass        &    time       & luminosity$^*$ \\
            &   RA.            Dec.      & [km~s$^{-1}$]  &   [km~s 
$^{-1}$]  &  [Mpc]    &    [Mpc]      &    [10$^{13}$ M$_\odot$]  &    [10$^{13}$ M$_\odot$]     & time$\times$ H$_0$    &      [10$^{11}$ L$_\odot$] \\
$FUV$ & 177.99642  16.802330 &3746$\pm$67&165$\pm$47& 49.94 &0.077$\pm$0.004&  0.23$\pm$0.09  &  0.77$\pm$0.25  & 0.04$\pm$0.02  &242$\pm$24.0\\
$NUV$ & 177.99938 16.804332 &3748$\pm$64 &157$\pm$44& 49.97 &0.071$\pm$0.005& 0.19$\pm$0.09 & 0.70$\pm$0.24& 0.04$\pm$0.02&1.46$ \pm$0.25\\
$B$   & 178.01486 16.820691 &3745$\pm$39 & 96$\pm$27& 49.93 &0.055$\pm$0.008& 0.06$\pm$0.03 & 0.25$\pm$0.09& 0.03$\pm$0.03 &0.14$\pm $0.02 \\
$r$   & 178.01821 16.824264 &3739$\pm$41&100$\pm$28& 49.85  &0.053$\pm$0.005&0.06$\pm$0.03   & 0.25$\pm$0.09& 0.03$\pm$0.03 &0.19$\pm $0.02\\

\hline
\end{tabular}

$^+$ Distance is derived from redshift adopting $H_0$=75 km sec$^{-1}$ Mpc$^{-1}$; $^*$ group luminosities refer to their respective band.

\end{table*}


\subsection{Comparing the data with models}

For the two brightest galaxies of this group (NGC 3933 and NGC 3934) we have extensive sets of data
from the Far-UV to the FIR, they are considered as a close pair in the FIR (Geller et al. 2006) and are rich in cold gas (Galletta et al. 1997). All these data put strong constraints on the evolution of these galaxies;  the dynamical analysis of the group is  another important constraint for our models. With the aim to match the overall SEDs and the global properties of these galaxies in a consistent way with the dynamical properties of the group, we select  two cases from a large set of SPH simulations which give us insights into the evolution of these systems and of the group itself.

Our SPH simulations of galaxy formation and evolution are starting from the same initial conditions as described in Mazzei \& Curir (2003; MC03 in the following) and Mazzei (2003) i.e., a  collapsing  triaxial systems initially composed of dark matter (DM) and gas in 
 different proportions and different total masses.
With respect to MC03, the particle resolution is enhanced here to 4$\times$10$^4$
 instead of  2$\times$10$^4$, so there are 2$\times$10$^4$ particle of 
 gas and 2$\times$10$^4$ of DM at the beginning in each new simulation.

Moreover, a set of simulations of galaxy encounters 
involving systems with 1:1 mass ratios and the  same initial 
conditions as in MC03 was also performed.
By seeking to exploit a wide range of orbital parameters, we 
carried out different simulations for each couple of interacting 
systems, varying the orbital initial conditions 
for to acquire the ideal Keplerian orbit of two equal point masses of mass 
equal to 10$^{12}$,  or 10$^{13}\,M_\odot$, the first peri-center 
separation,  $p$, equal to the initial length of the major axis of the 
dark matter triaxial halo, i.e. 88 kpc for 10$^{12}\, M_\odot$, or 
equal to 1/10, 1/7,  and 1/5 of the same axis for 10$^{13}\,M_\odot$ 
encounters. 
For each of these separations, we changed the eccentricity to obtain hyperbolic orbits of different energy. The spins 
of the systems are generally parallel to each other and perpendicular 
to the orbital plane, so we studied direct encounters however, 
some cases with misaligned  spins were also analyzed  to deepen the effects of the system initial rotation on 
the results. Moreover, for a given set of encounters with the 
same orbital parameters  we also examined the role of 
increasing initial gas fractions. These simulations will be 
fully discussed in a different paper (Mazzei, 2011 in prep).

All simulations  include self--gravity of gas, stars and DM, 
radiative cooling, hydrodynamical pressure, shock heating, 
artificial viscosity, star formation (SF) and  feedback from 
evolved stars and type-II SNe, as in MC03.

The initial mass function (IMF) is of Salpeter type with 
upper mass limit 100$\,M_\odot$ and lower mass limit 0.01$\,M_\odot$ (Salpeter 1955); (see 
MC03 and references therein for a discussion).
Indeed all our simulations provide the synthetic SED 
at each evolutionary step. The SED accounted for chemical 
evolution, stellar  emission, internal extinction and 
re-emission by dust in a self-consistent way, 
as described in (Spavone et al. 2009 and references therein); 
this extends over four orders of magnitude in wavelength,
 i.e., from 0.1 to 1000 $\mu$m. So, each simulation self-consistently 
 provides morphological, dynamic and chemo-photometric evolution. 
 
Two simulations, able to match the global properties 
 of NGC~3934 and NGC~3933, are chosen and  discussed in detail below.
These point toward an independent evolution of NGC 3933 
 whereas NGC 3934, is the result of a major merger, whose final phase began 3 Gyr ago.

\begin{figure}
  \centering
 {\includegraphics[width=9cm]{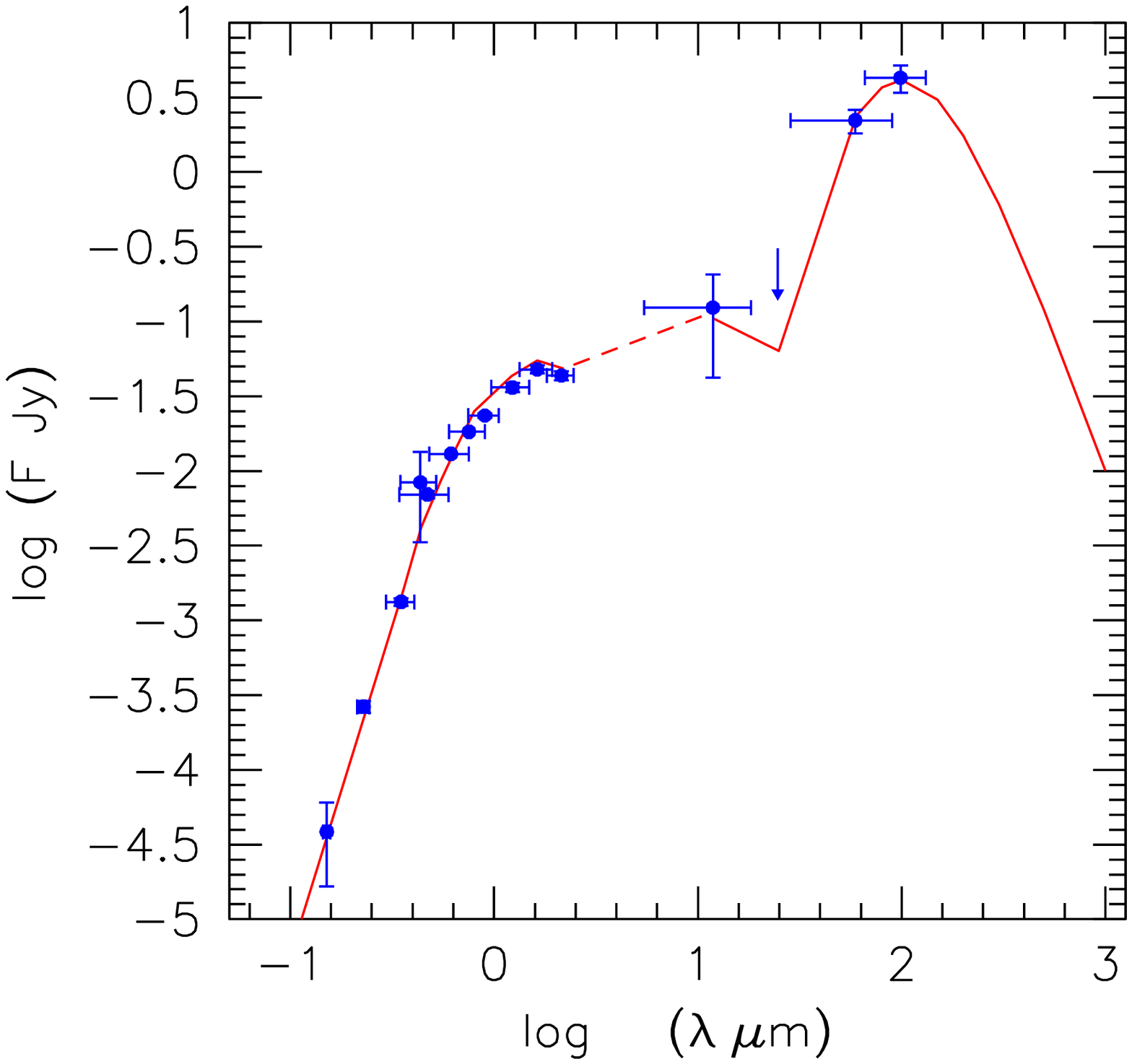}
  \includegraphics[width=9cm]{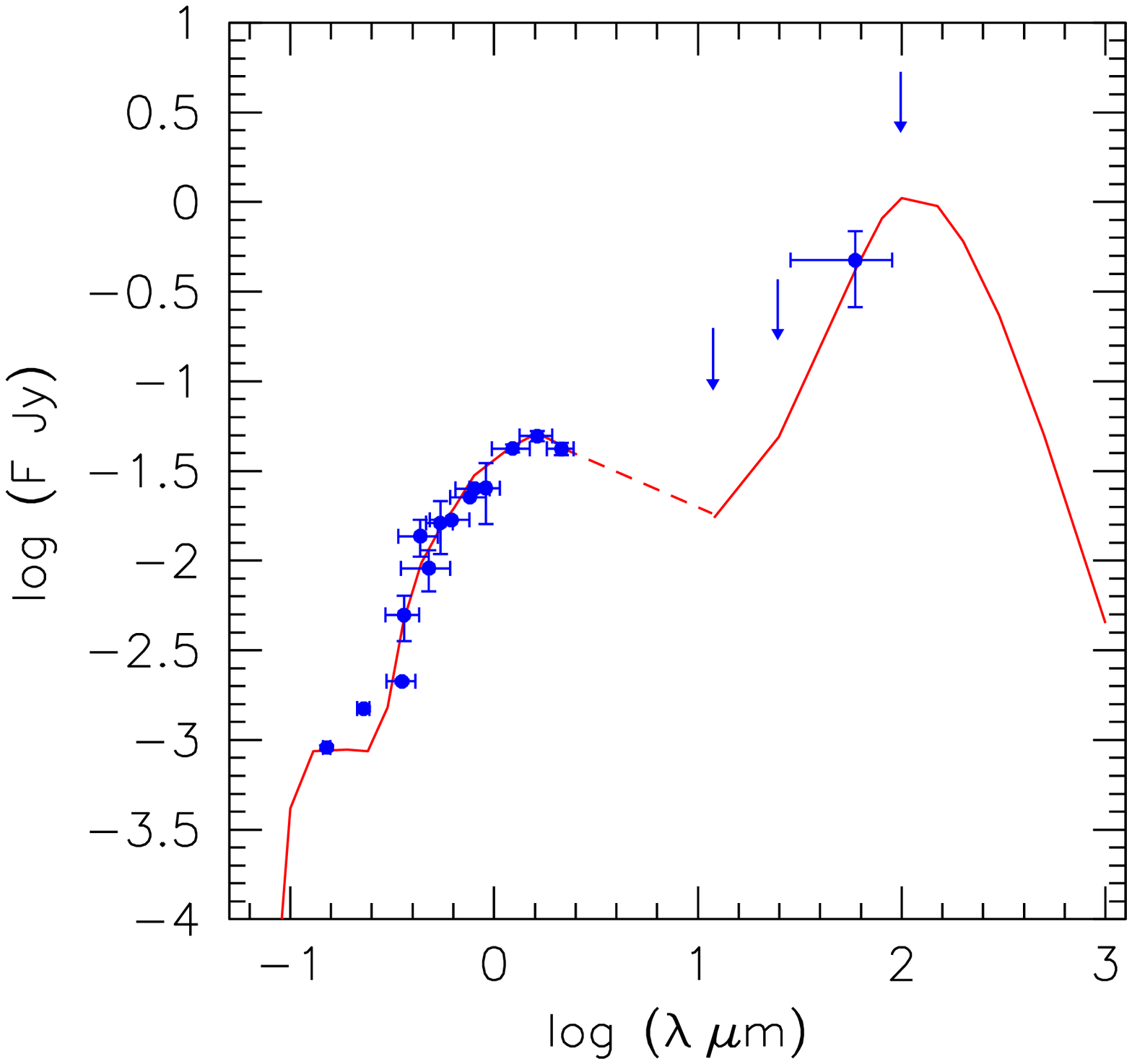}}
      \caption{ {\sl{Top}}: continuous line (red) shows the prediction of our model (see
      text) for NGC~3934. (Blue) filled circles correspond to data from Table~\ref{Table2} for FUV and NUV, from Table~\ref{Table3} for the B- and R- and from NED for the remaining bands. Arrows show upper limits; error bars account 
      for band width and 3$\sigma$ uncertainties.
	{\sl{Bottom}}: symbols are the same as in the top panel, but for NGC~3933. }
       \label{Fig_sed}
   \end{figure}

\subsubsection{NGC~3933}

We derive an adequate match of dynamical and
photometric properties of NGC~3933  with a simulation 
of total mass of 10$\times10^{11}$ M$_\odot$.
The starting point  is  a triaxial collapsing system initially composed of 4$\times$10$^4$ particles,
2$\times$10$^4$ of gas and 2$\times$10$^4$ of DM 
with a relative mass ratio 0.1. The  system is built up as described in
MC03, i.e., with a spin parameter, $\lambda$, given by 
$|{\bf {J}}||E|^{0.5}/(GM^{0.5})$, where E is the total energy, J the total angular momentum and G the gravitational
constant, equal to 0.06 and aligned with the shorter principal axis of the DM
halo; the triviality ratio of the DM halo, $\tau=(a^2-b^2)/(a^2-c^2)$ 
(Warren et al. 1992), where $a>b>c$, is 0.84 with an average radius of about 500 kpc.

Fig.  \ref{Fig_sed} (bottom panel) compares the predicted SED with the
available data, accounting for inclination between line  
of sight and polar axis, of 74.4 degrees ({\it HYPERLEDA}; Mazzei \& De Zotti 1992, Bekki \& Shioya 2001).

\begin{figure}
  \centering
 {\includegraphics[width=9cm]{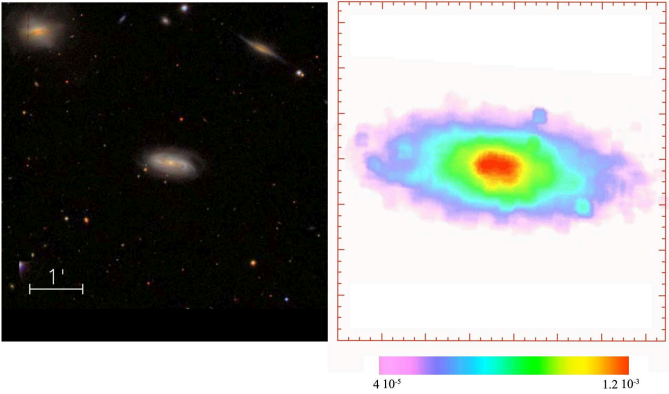}} 
      \caption{Color-composite SDSS image of NGC~3933 group {\sl (left)} 
      centered on the spiral galaxy  NGC~3933 is compared on the same scale 
      with the R- band map  from our simulation {\sl (right)}; the map includes 
      forty equi-spaced levels and a density contrast of three hundred 
normalized to the total flux in the map. Note that the observed galaxy (left panel) 
corresponds to the inner (red) region in the right panel.}
       \label{Fig_sim1}
   \end{figure}

Dynamical predictions, in particular the maximum gas rotation 
velocity  of 130 km~s$^{-1}$, agree with kinematic features observed  
for NGC~3933,  (136.8$\pm$15.2 km~s$^{-1}$, {\it HYPERLEDA}).
The predicted  average age of stellar populations 
is  4.5 Gyr and the total star formation rate (SFR), 2~M$_\odot$~yr$^{-1}$, is consistent with the value derived from the L$_{1.4GHz}$ (Sec 3.2). This SFR is higher than expected on average for its morphological type, S0a (Kennicutt 1998),  it is more similar to the average total SFR of the Milky Way (Yin et al. 2009). The M/L$_R$ ratio of stars is about 3  $M_\odot/L_\odot$ at the optical radius (i.e, R$_{25} \simeq$ 9 Kpc  using data from  {\it NED}\footnote{ (http://nedwww.ipac.caltech.edu/)})
and  the stellar mass  1.2$\times 10^{10}\, M_\odot$, 
about half of the total mass at the same radius. 
Figure \ref{Fig_sim1} shows the R- band simulated map of this snapshot.
The mass of gas with temperature lower than 10$^4$\,K, which 
represents the upper limit of the cold gas mass in the system 
(its cooling time scale is much shorter than the snapshot time range, 0.154 Gyr),
is 3.5$\times 10^9\, M_\odot$,  which agrees well with the results 
of Galletta et al (1997). 
The bolometric luminosity of the whole galaxy is 3.21$\times 10^{10}\,L_\odot$. 
The total mass inside 55 Kpc is about 3$\times 10^{11}\,M_\odot$, rising to 3.6$\times 10^{11}\,M_\odot$ within 77 Kpc.
 
The far-IR  (FIR) SED is composed of a warm and a cold dust 
component and includes PAH molecules (Mazzei \& De Zotti 1992). 
Warm dust is located in regions of the high-radiation field,  i.e., in the 
neighborhood of OB clusters, whereas cold dust is heated by the 
general interstellar radiation field. 
The energy ratio  between the warm and the cold dust components in 
Fig. \ref{Fig_sed} (bottom panel) is 0.3, the intensity of diffuse radiation 
field is about three times higer than the value used to match the Galactic FIR emission 
Mazzei \& De Zotti (1992) and the warm dust temperature is about 67~K.

\subsubsection {NGC~3934}

\begin{figure}
  \centering
 {\includegraphics[width=9cm]{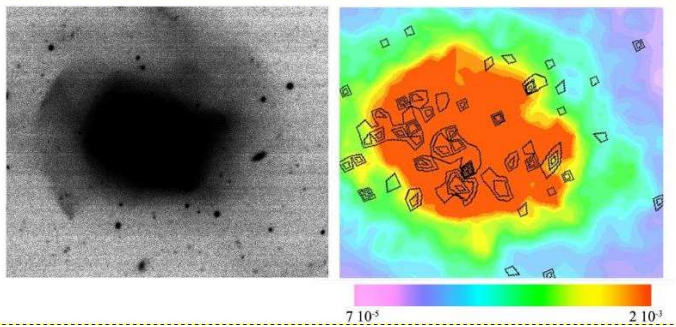}}
      \caption{ Deep R- band image of NGC~3934 {\sl (left)} compared on the
       same scale with isodensity contours {\sl (right)} from our simulation (see text). 
       The simulated image includes twenty-six no-equispaced levels to emphasize 
       the shells: seven  equispaced levels from 7$\times 10^{-5}$ to 
       1 $\times 10^{-4}$, eighteen equispaced levels from 1.5$\times 10^{-4}$ to 
       1 $\times 10^{-3}$ and then the last level, 2 $\times 10^{-3}$ normalized 
       to the total counts in the map.  Black contours in the right panel correspond 
       to star clusters 0.05 Gyr old or younger.
}
       \label{Fig_sim2}
   \end{figure}

The whole SED and the global properties of NGC~3934 are matched well by a  merger simulation 
with initial systems of equal total mass, 10$^{13}$ M$_\odot$, and a gas 
fraction of 0.01; the first peri-center separation is 175~kpc, 
corresponding to 1/10 of the major axis, and the  eccentricity 1.2.
Initial star systems born in the inner regions of their halos after 2.5 Gyr 
from the beginning grow, changing their shapes step by step as their 
trajectories are approaching.
Stellar systems merge in a unique configuration after 9 Gyr. However, 
the merger event does not complete until their centers of mass lose all 
their residual energy to coalesce in the center of mass of the whole 
system. In the meantime a shell system arises from their oscillations 
which on average expands and rarefies gradually, showing also 
re-feeding phases owing to the complex motions, shocks, viscosity 
and friction of several system components involved in the process of merging.
The top panel of Fig.  \ref{Fig_sed} compares the predicted SED with the
available data accounting for 45 degrees of inclination between line of sight 
and polar axis ({\it HYPERLEDA}).
The SED corresponds to the same age as NGC~3933, i.e. 
to 12.3 Gyr. However, the average age of stellar populations  is younger than in the section above, 
3-4 Gyr, with a higher  average total SFR,  about 5 M$_\odot$~yr$^{-1}$, which 
remarkably agrees with the SFR we derived from the continuum radio emission at 1.4GHz.
 Gas and stars draw complex structures: ripples arise if one looks at the system 
 morphology in different projections.
The FIR SED requires an average  temperature  of warm dust 
equal to 45K, a cold dust radiation field with intensity about 
half than that used to match the FIR SED of NGC~3933,  but a higher
 warm-to-cold energy ratio, of 0.5.\\
\indent
Figure \ref{Fig_sim2} compares on the same scale a deep R- band image of
NGC~3934 with isodensity contours of our simulation. Contours of  star clusters younger than 0.05 Gyr are overplotted. They match the region of UV emission well, i.e. both the main body of the galaxy and the shell regions, as observed in Fig.\ref{Fig_UV} (left panel).

The bolometric luminosity of the whole galaxy  is about twice that of NGC~3933.
The total mass inside 55 Kpc is about 7$\times 10^{11}\,M_\odot$ and 
11$\times 10^{11}\,M_\odot$  inside 77 Kpc  which agrees well with the mass of 13$\times 10^{11}\,M_\odot$ obtained
from our measured velocity dispersion (see Section 2.1) using the relation M=5$\times\sigma^{2}R_e/G$ (Bertin et al. 2002).

The dynamical results reported in Table~4 are compatible with the estimates
of masses and radii provided by the above simulations of NGC~3933 and NGC 3934.  

 
\section{Summary and conclusions}

We analyzed the multi-wavelength properties of NGC 3934 and its environment.
The multi-band photometric analysis permits us to study in detail the morphology and the SED of NGC~3934,
 the second brighter member of the group, and to investigate the nature 
 of this galaxy which is supposed to be a polar.ring galaxy  (Whitmore eet al 1990). 

Our study shows that NGC 3934 is not a polar.ring galaxy, as suggested by the literature.
The peculiar morphology of this galaxy is caused by a prominent dust-lane structure and by a wide type-II shell structure.
The shell system together with the presence of the strong dust--lane suggest that NGC 3934 is the product of  a ``wet'' merging episode. This is confirmed also by the presence of some NUV knots along the NW ripple of NGC 3934 
 similar to those detected by GALEX in another shell galaxy, NGC 1210. These features are the debris of a  recent accretion event that is responsible for the shell structure (Marino et al. 2009).

Our spectroscopic study of PGC~213894 allows us to add this galaxy to the bright members of 
 the group whose projected core is composed of five galaxies, with  Hickson's 
 compact group characteristics.

Our multi-band photometric data allow us to perform a dynamical analysis of the group (Table ~4). The relatively short crossing times indicate that the group is probably virialized. The low
velocity dispersion could either argue against the virialization of the group
(chance alignment with a loose group) or suggest that the tidal friction 
is slowing down the galaxies and which brings the group toward the full coalescence.   
The virial masses of the group are small, lower than 2.3$\pm$0.9 $\times10^{12}$ $M_\odot$. 

We performed and analyzed a large set of SPH simulations to 
match the morphology and the global properties of the brighter 
optical members of the group, namely NGC~3933 and NGC~3934, and to give insights into the group evolution. The simulation matching NGC~3933, a typical spiral in the group, is consistent with an unperturbed evolution of this system with an SFR similar to that of the Milky Way.   On the other hand NGC 3934 appears to be
the result of a major merger (see also Weil \& Hernquist 1993), whose final phase began 3 Gyr ago.  Our simulation points at a "wet' merger, i.e. a merger where dissipation plays a meaningful role on the evolution,
as observations suggest. Moreover, the simulation performed is able to account for both the whole SED of  this galaxy from the UV to FIR spectral range, and for the morphology, including the shell structure and the distribution of the UV emission.

 Since their discovery (Malin \& Carter 1983), shell galaxies are noticed
 to inhabit low-density environments, specially avoiding clusters. One 
 remarkable  exception is NGC~4552 in Virgo.  NGC~3934 could
 represent an example of  a shell galaxy in a dense (at least in projection) galaxy 
 association. The presence of NGC~3934 makes this group an 
 evolving galaxy association, probably at an early stage of its evolution, as
 suggested by the nearly unperturbed morphologically and photometric nature 
 of its spiral members.  
   
\begin{acknowledgements}

We tank the anonymous referee for useful comments.
We acknowledge the financial contribution from the agreement ASI-INAF I/009/10/0.
 {\it GALEX} is a NASA Small Explorer, launched in April
2003. {\it GALEX} is operated for NASA by California Institute of
Technology under NASA contract  NAS-98034. 
This work is based on {\it GALEX} data from GI program 59 and 
archival data. The data were obtained from MAST (http://galex.stsci.edu).
This research has made
use of the SAOImage DS9, developed by Smithsonian Astrophysical 
Observatory and of the NASA/IPAC Extragalactic Database ({\tt NED}) which 
is operated by the Jet Propulsion Laboratory, California Institute of
Technology, under contract with the National Aeronautics and Space
Administration.  
We acknowledge the usage of the {\tt HYPERLEDA} database (http://leda.univ-lyon1.fr).
Funding for the SDSS and SDSS-II has been provided by the Alfred P. 
Sloan Foundation, the Participating Institutions, the National Science 
Foundation, the U.S. Department of Energy, the National Aeronautics and Space 
Administration, the Japanese Monbukagakusho, the Max Planck Society, and the 
Higher Education Funding Council for England. The SDSS Web Site is http://www.sdss.org/.
The SDSS is managed by the Astrophysical Research Consortium for the Participating Institutions. 
The Participating Institutions are the American Museum of Natural History, Astrophysical Institute 
Potsdam, University of Basel, University of Cambridge, Case Western Reserve University, University 
of Chicago, Drexel University, Fermilab, the Institute for Advanced Study, the Japan Participation 
Group, Johns Hopkins University, the Joint Institute for Nuclear Astrophysics, the Kavli Institute 
for Particle Astrophysics and Cosmology, the Korean Scientist Group, the Chinese Academy of Sciences 
(LAMOST), Los Alamos National Laboratory, the Max-Planck-Institute for Astronomy (MPIA), the 
Max-Planck-Institute for Astrophysics (MPA), New Mexico State University, Ohio State University, 
University of Pittsburgh, University of Portsmouth, Princeton University, the United States Naval Observatory, and the University of Washington. We tanks the anonymous referee for useful comments.
\end{acknowledgements}

\end{document}